\documentclass[fleqn,usenatbib]{mnras}

\usepackage{newtxtext,newtxmath}
\usepackage[T1]{fontenc}

\DeclareRobustCommand{\VAN}[3]{#2}
\let\VANthebibliography\thebibliography
\def\thebibliography{\DeclareRobustCommand{\VAN}[3]{##3}\VANthebibliography}

\usepackage{graphicx}	% Including figure files
\usepackage{amsmath}	% Advanced maths commands
\usepackage{longtable}
\usepackage{verbatim}
\usepackage{array,multirow,makecell}
\usepackage{rotating}
\usepackage{dirtytalk}
\usepackage{tablefootnote}
\usepackage{float}
\usepackage[version=4]{mhchem}
\usepackage{silence}
\usepackage{tablefootnote}
\usepackage[flushleft]{threeparttable}
\newcommand{\pcms}{cm$^{-2}$}
\newcommand{\pcmc}{cm$^{-3}$}
\newcommand{\kms}{km s$^{-1}$}
\newcommand{\s}{s$^{-1}$}
\title[An intriguing chemically active outflow in OMC-2/3]{Discovery of an intriguing chemically rich outflow in the OMC-2/3 filament}

\author[M. Bouvier et al.]{M. Bouvier,$^{1,2}$
\thanks{E-mail: bouvier@strw.leidenuniv.nl}
L. Giani,$^{2}$
L. Chahine,$^{2}$
A. L\'opez-Sepulcre,$^{2,3}$
C.Ceccarelli,$^{2}$
L. Podio$^{4}$
\\
$^{1}$Leiden Observatory, Leiden University, P.O. Box 9513, 2300 RA, Leiden, The Netherlands\\
$^{2}$Univ. Grenoble Alpes, CNRS, IPAG, 38000 Grenoble, France\\
$^{3}$Institut de Radioastronomie Millimétrique (IRAM), 300 rue de la Piscine, 38400 Saint-Martin d’Hères, France\\
$^{4}$INAF, Osservatorio Astrofisico di Arcetri, Largo E. Fermi 5, I-50125, Firenze, Italy
}

\date{Accepted XXX. Received YYY; in original form ZZZ}

\pubyear{2024}

\begin{document}
\label{firstpage}
\pagerange{\pageref{firstpage}--\pageref{lastpage}}
\maketitle

% Abstract of the paper
\begin{abstract} %250 words
Studying chemically rich protostellar outflows and their jet provides an important insight into the low-mass star formation process and its related chemistry. Whilst well-known shock tracers such as SiO can be used to study the jet properties and give information about the dynamics of the system, interstellar complex organic molecules (iCOMs) have been useful in constraining the age of shocked gas, for example. Yet, the number of outflows mapped in iCOMs is still limited. In this work, we study the outflow driven by the protostar FIR6c-a (HOPS 409) located in the OMC-2/3 filament. We report the detection of the red-shifted jet, left undetected in previous studies, as well as the detection of the iCOMs methanol (CH$_3$OH) and methyl cyanide (CH$_3$CN) for the first time towards this outflow.
Using SiO, we derived some jet properties (i.e., collimation and dynamical time). We found a clear dichotomy between the blue- and red-sifted jets, likely due to the density of the medium in which the jets propagate. In addition, we identified two bow shocks within the blue-shifted part of the outflow, which we attribute to two different ejection events. Finally, using the \ce{CH3OH} and \ce{CH3CN} abundance ratio and chemical modelling, we constrained the outflow age to be $\geq 1000$ yr old and, surprisingly, found that a cosmic-ray ionization rate of $10^{-14}$ \s~ is needed to reproduce the observed ratio towards the source.
\end{abstract}

% Select between one and six entries from the list of approved keywords.
% Don't make up new ones.

%%%%%%%%%%%%%%%%%%%%%%%%%%%%%%%%%%%%%%%%%%%%%%%%%%

%%%%%%%%%%%%%%%%% BODY OF PAPER %%%%%%%%%%%%%%%%%%

\section{Introduction} \label{sec:intro}

Within the low-mass star formation process, the protostellar phase is where most of the accretion from the surroundings onto the central forming core occurs. In return, a supersonic ($\sim 50-500$ \kms) and collimated jet is ejected from the protostellar/disk system \citep[e.g.,][and references therein]{bally2007,frank2014}. The launched jet accelerates the surrounding gas, leading to the formation of a molecular outflow that can spread at velocities between $\sim 10-100$ \kms from hundreds of au to a few parsecs \citep[e.g.,][and references therein]{arce2007, bally2007}. The shocks induced by the jet propagation are also thought to be possible acceleration sites for cosmic rays (CR; e.g. \citealt{padovani2015, padovani2016, GO18}), drivers of complex chemistry in molecular clouds \citep[e.g.][]{padovani2009a, grenier2015}.  
The most powerful outflows are driven by the youngest embedded protostars (Class 0 protostars; \citealt{bontemps1996, saraceno1996}), which are thus ideal laboratories where to study their physical and chemical features.

Protostellar outflows are typically observable in the millimetre via low-J transition of CO lines \citep[e.g.,][]{lada1985, bontemps1996,saraceno1996, williams2003, AS06, jorgensen2007, dunham2014, tobin2016b, feddersen2020}. 
However, a few of them have been characterised as \textit{chemically active} outflows as they show the abundance enhancement of several other molecular species (e.g., SiO, \ce{CH3OH}, H$_2$O, S-bearing species) due to endothermic chemical reactions and ice mantle sputtering or grain shuttering \citep[see e.g.,][]{Hollenbach1989, Flower1994, Draine1995}. 
Such outflows are particularly interesting to study as they offer a unique opportunity to investigate shock chemistry and the complex chemical reactions involved \citep{TB11, codella2017, giani2023, chahine2024}.

Among the species enhanced in these shocks, SiO, a well-known shock tracer \citep[e.g.,][]{schilke1997b, caselli1997, gusdorf2008b},  typically traces the high-velocity jet \citep[e.g.,][]{hirano2006, cabrit2007, cabrit2012, codella2007,lopez2011, codella2014a, lee2007a,lee2007b, tafalla2015, podio2016, podio2021, bjerkeli2019, chahine2022b}. If the angular resolution is high enough, the knotty structures caused by the episodic nature of the jet \citep[e.g.,][]{raga1990, SN93} can be resolved and used to derive the jet properties (e.g., collimation angle, timescale of the episodic ejections; \citealt{plunkett2015, matsushita2019, lee2020, dutta2024, jhan2022, takahashi2024}). 

Methanol, \ce{CH3OH}, known to form on the icy grain mantles \citep{WK02, rimola2018}, is the simplest and more widely detected interstellar Complex Organic Molecules (iCOM; \citealt{herbst2009, ceccarelli2023}) observed in protostellar shocks \citep[e.g.,][]{bachiller1995,gibb1998, garay2002, codella2010, sakai2012, lefloch2017, holdship2019b, desimone2020b, vastel2022}. 
More complex species have also been detected towards protostellar outflows, from low-mass \citep[e.g.,][]{arce2008, oberg2011, lefloch2017, mendoza2014, codella2017, codella2020, holdship2019b, desimone2020b}, to intermediate- and high-mass protostars \citep[e.g.][]{liu2002, favre2011, leurini2011, palau2017, rojas-garcia2022, rojas-garcia2024, busch2024}. In some of these studies, the iCOMs abundance ratio with respect to methanol has been used to study the iCOM formation mechanisms \citep{codella2017, codella2020, burkhardt2019,  desimone2020b} or to constrain the shock age ([\ce{CH3OH}]/[\ce{CH3CN}]; \citealt{giani2023}).  

\begin{table*}
    \centering
       \caption{Detected species and transition lines, their parameters, and channel spacing and primary beam size of the associated spectral windows.}
    \label{tab:species-spw}
    \begin{tabular}{cccccccccc}
    \hline \hline
         Molecule & Frequency & Transition & $E_{\text{up}}$& $g_{\text{up}}$ & $A_{\text{ij}}$  & beam  & primary beam & rms  \\
         &[MHz] & & [K] & & [$\times 10^{-5}$\s] &[$\theta_{\text{maj}}'' \times \theta_{\text{min}} '' \text{PA}(^\circ)$] &[$''$]& [mJy.beam$^{-1}$]\\
         \hline
         \ce{CH3OH} &218440&$4_{-2,3} - 3_{-1,2} $ E& 45.5 & 36 & 4.69 & $0.52 \times 0.29 \ (107)$ & 28.8&3.5 \\
         &243915 &$5_{1,4} - 4_{1,3} $ A& 49.7 &44 & 5.97 &$0.32 \times 0.28 \ (-78)$ &25.8&2.6\\
         &261805 &$2_{1,1} - 1_{0,1} $ E& 28.0 &20 & 5.57  &$0.30 \times 0.25 \ (-77)$ &24.1&2.7\\
         \hline
         SiO & 260518& $6 - 5$ &43.8 & 14 & 91.2 &$0.30 \times 0.26 \ (104)$ &24.2& 2.6 \\
         \hline
          \ce{CH3CN}& 257507 &$14_2-13_2$ & 121.3&3 &144.5&\multirow{3}{*}{$0.31 \times 0.26 \ (107)$} &\multirow{3}{*}{24.5}&\multirow{3}{*}{3.0} \\
         &257522 &$14_1-13_1$ &99.8 &3&146.9&\\
         &257527& $14_0-13_0$& 92.7&3 &147.6& &  \\
         \hline
    \end{tabular}
    \begin{tablenotes}
    \item [] Frequencies and spectroscopic parameters have been extracted from the CDMS catalogue \citep{muller2005}. For \ce{CH3OH} (TAG 032504, version 3$^*$), SiO (TAG 044505, version 2$^*$), and \ce{CH3CN} (TAG 041505, version 2$^*$), the available data are from \citet{xu2008}, \citet{muuller2013}, and  \citet{muuller2015}, respectively.\end{tablenotes}
 %\tablecomments{}
\end{table*}

However, to either derive the jet properties, constrain the iCOM formation route or the shock age requires sub-arc-second angular resolutions. Whilst recent efforts have been made to increase such studies of protostellar jets \citep[e.g.][]{tychoniec2019, podio2021, jhan2022, dutta2024}, the number of low-mass protostellar outflows mapped in iCOMs is still very limited \citep[e.g.][]{codella2017, codella2020, desimone2020b, sabatini2024}. To understand the Class 0 outflow process and the origin of its associated chemical richness, more high-angular studies of chemically active outflows are needed. In this work, we present the discovery of a chemically active protostellar outflow located in the Orion Molecular Cloud (OMC) 2/3 filament.

Fir6c-a (also known as HOPS-409; \citealt{furlan2016}) is a low-mass Class 0 protostar located in the OMC-2/3 filament within the Orion A molecular cloud at a distance of (393 $\pm$ 25) pc \citep{grosschedl2018}. The bolometric luminosity of the source is 8.2 $L_{\odot}$ \citep{furlan2016}. The protostar is known to drive one of the most extended bipolar outflows within the filament.
The \ce{H2} flow detected by \citet{stanke2002} is even extending at a length of $\sim 0.41$ pc whilst the CO(1-0) emission shows a northern blue-shifted lobe longer than 0.2pc, with an inclination of 50$^{\circ}$ \citep{tobin2016b}. The north part of the outflow has been observed in several CO transitions (both in the millimetre and the infrared), \ce{H2O}, and in SiO (2-1) and (5-4) transitions \citep{shimajiri2009, tobin2016b, gomezruiz2019, tanabe2019}. The SiO emission traces a well-collimated jet in the northern blue-shifted lobe \citep{shimajiri2009, gomezruiz2019}. \citet{shimajiri2009} found the SiO emission to be composed of two components which are misaligned, suggesting that the ejections were done in different 3D directions (as initially proposed by \citealt{reipurth1996}). The red-shifted part of the jet has not been detected. Finally, FIR6c-a was part of the sources sampled by the ORion ALMA New GEneration Survey (ORANGES; \citealt{bouvier2021, bouvier2022}) and was found to be one of the few sources hosting a hot corino \citep{bouvier2022}. Whilst investigating the emission of \ce{CH3OH}, the most abundant iCOM in hot corinos, we found that \ce{CH3OH} was also detected towards the outflow. The FIR6c-a outflow thus provides an ideal chemically active outflow to be studied in the region of Orion.

In this work, we present the first high-angular resolution observations (0.25$''$ or 100 au) of the outflow driven by the FIR6c-a protostar. 
The paper is structured as follows.
The observations are presented in Sect.~\ref{sec:obs}. 
In Sect.~\ref{sec:emission-distribution}, we present the emission distribution maps, in Sect.~\ref{sec:conditions} we extract the physical conditions of the emitting gas, in Sec.~\ref{sec:jet} we give the characteristics of the jet and in Sect.~\ref{sec:model} we present the astrochemical modeling. 
We discuss our findings in Sect.~\ref{sec:disc} before summarising our work in Sect.~\ref{sec:conclusions}.

%----------------------------------------------------------

\section{Observations} \label{sec:obs}

The observations were performed between 2016 October 25th and 2017 May 5th during Cycle 4, under the ALMA project 2016.1.00376.S (P.I. Ana L\'opez-Sepulcre). The phase centre of the observations was set to $\alpha$(J2000)$=05^h35^m21.60^s$ and $\delta$(J2000)$=-05^\circ 13'14.00''$ and at a systemic velocity of $V_{lsr}=10.80$ \kms. Two different spectral setups, within the spectral ranges  $218-234$ and $243-262$ GHz, were used and centred on various molecular species. The narrow spectral windows have a bandwidth of 58.59 MHz with a channel spacing of 122 kHz ($\sim$ 0.15-0.17 \kms) whilst the wide spectral windows have a bandwidth of 1875 MHz with a channel spacing of 0.977 MHz ($\sim 1.2-1.3 $\kms). The bandpass and flux calibrators used were J0510+1800 and J0522-3627, and J0607-0834 and J0501-0159 were used for the phase calibration. The precipitable water vapor (PWV) was, on average, less than 1mm, and the phase rms noise smaller than 60$^\circ$. The flux calibration error is estimated to be better than 10\%. 

\begin{figure*}
    \centering
    \includegraphics[width=0.9\textwidth]{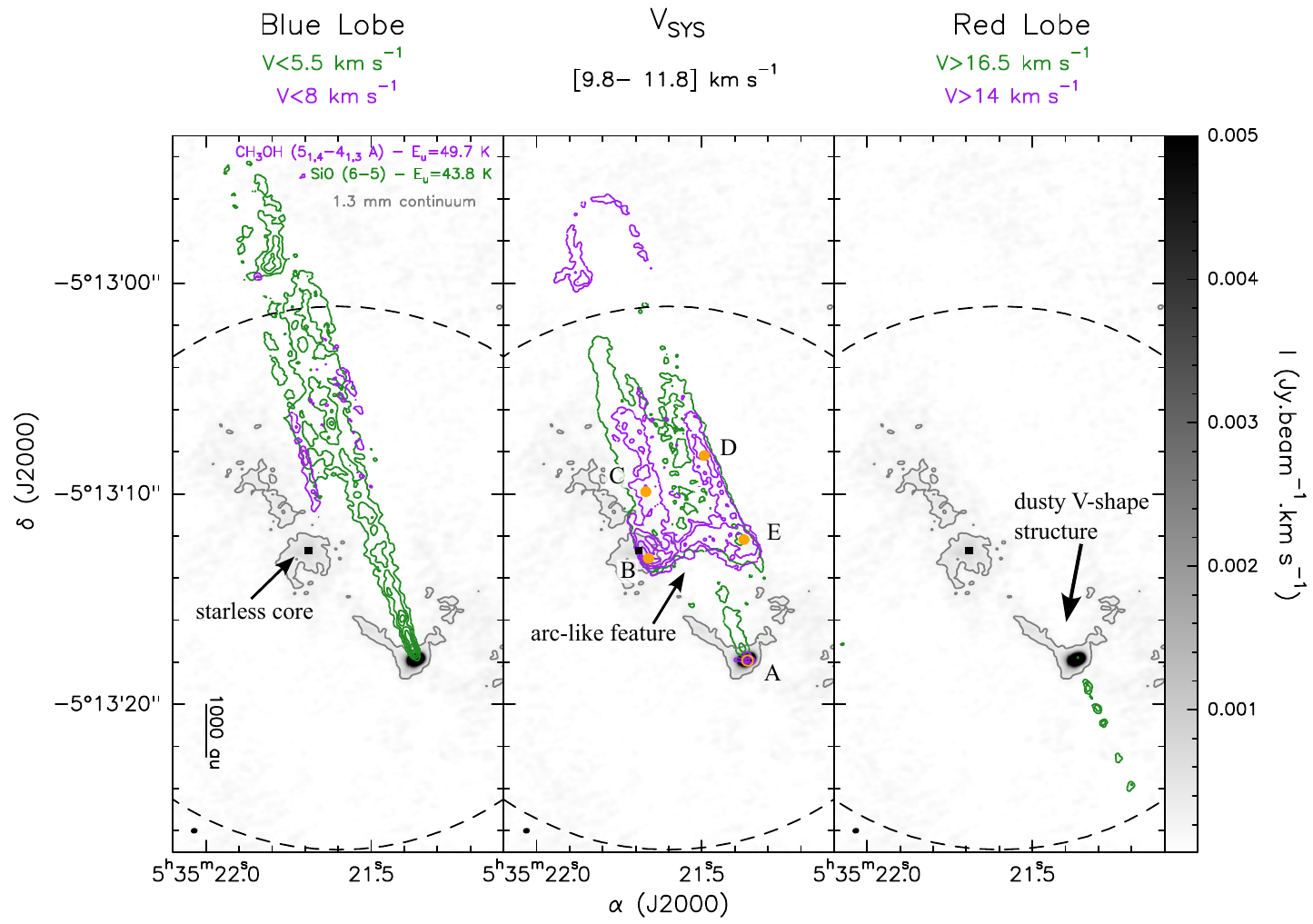}
    \caption{Moment 0 maps of SiO (6-5; green contours) and \ce{CH3OH} ($5_{1,4}-4_{1,3}$ A; purple contours), overlaid on the 1.3mm continuum emission (grey-scaled background). For the 1.3mm continuum, only the 3$\sigma$ level is shown (grey contour; 1$\sigma=60 \mu$Jy.beam$^{-1}$\kms.) The moment maps are shown at three different velocity components: blue-shifted ($V<5.5$ and $V<8$ \kms for SiO and \ce{CH3OH}, respectively; \textit{left}), around the systemic velocity ([$10.8 \pm 1$] \kms; \textit{centre}), and red-shifted ($V>16.5$ and $V>14$ \kms for SiO and \ce{CH3OH}, respectively; \textit{right}). Levels of SiO start at 10, 5, and 10 $\sigma$ ($1\sigma=2.6$ mJy.beam$^{-1}$\kms) with steps of 40, 15, and 15 $\sigma$, respectively. Levels of \ce{CH3OH} start at 4 $\sigma$ ($1\sigma=2.6$ mJy.beam$^{-1}$\kms) with steps of 7$\sigma$. The black filled square mark the position of the starless core FIR6c-c \citep{kainunlainen2017, bouvier2021}. The (filled) orange circles show the regions (A to E) of extraction of the spectra}. The dashed black circle represents the primary beam of 25.8$\arcsec$ from the \ce{CH3OH} observation. The beam from the SiO observation is depicted in the bottom left corner of each panel. The red-shifted part of the SiO jet is detected for the first time. 
    \label{fig:moment-maps}
\end{figure*}

Data calibration was first performed using the Common Astronomy Software Application (CASA; \citealt{mccmullin2007}) before performing the imaging in GILDAS\footnote{http://www.iram.fr/IRAMFR/GILDAS} using the program MAPPING. Details about the continuum subtraction and phase self-calibration performed and applied to the continuum-free cubes are given in \cite{bouvier2021}. The cubes were re-sampled to a channel spacing of 0.5 \kms and were cleaned using a natural weighting (with the CLEAN procedure). The rest frequency of the detected lines toward the OMC-2 FIR6c-a outflow, and the associated beam, primary beam sizes, and the root-mean-square (rms) are summarised in Table \ref{tab:species-spw}.

%------------------------------------------------------------

\section{Emission distribution} \label{sec:emission-distribution}

We detect several species towards the FIR6c-a outflow, among which \ce{CH3OH}, SiO and \ce{CH3CN}. As we aim to characterise the jet and constrain the age of the outflow, we focus the present study on these three species. Another work dedicated to the chemical richness of the outflow will be performed at a later stage. The detected transitions of \ce{CH3OH}, \ce{CH3CN}, and SiO as well as their spectroscopic parameters are presented in Tab.~\ref{tab:species-spw}.

As already mentioned in the introduction, SiO had already been detected towards the jet of the FIR6c-a outflow at a larger scale \citep[e.g.][]{shimajiri2009, gomezruiz2019}, and \ce{CH3OH} was detected towards the protostar itself, tracing the hot corino \citep{bouvier2022}. In this study, we detect \ce{CH3OH} and \ce{CH3CN} towards the FIR6c-a outflow for the first time, which makes FIR6c-a the second iCOM-rich outflow of the OMC-2/3 filament, after HOPS 87 \citep{hsu2024}. The frequency range covered by the observations encompasses several methanol transitions, with upper-level energies ($E_{\text{up}}$) ranging from 28 to 190.4 K. However, only the transitions with $E_{\text{up}}$ ranging between 28 and 50 K are detected towards the outflow. For \ce{CH3CN}, we do not resolve the hyper-fine transitions of each K-ladder. We describe below the emission distribution of each of these detected species.

Figure \ref{fig:moment-maps} shows the spatial distribution of SiO (6-5) and \ce{CH3OH} ($5_{1,4}-4_{1,3}$; $E_u=49.7$ K) towards the FIR6c-a outflow overlaid on the 1.3mm dust continuum emission. The figure  \ref{fig:moment-maps} contains three panels, associated with the blue- and red-shifted part of the outflow (left and right panels, respectively) and with the emission around the systemic velocity of the source of $+$10.8 \kms (Central panel). For the blue-shifted component, we integrated the SiO and \ce{CH3OH} emissions between [-12.3 ; +5.5] and [-3; +8] \kms, respectively. The two ranges of integration are different as the blue-shifted emission covers a larger range of velocities in the case of SiO. In addition, since the linewidths of SiO are significantly broader than that of \ce{CH3OH} (see Fig.~\ref{fig:full_lines-profiles}), we defined the range to avoid any contamination from emission occurring at the rest velocity of the source ($V_{lsr}=10.8$ \ \kms). Around the $V_{\text{sys}}$, we integrated both SiO and \ce{CH3OH} in the range [+9.8 ; +11.8] \kms. Finally, we integrated the SiO emission in the range [+16.5 ; +26.7] \kms for the red-shifted component. Overall, the SiO emission in the Figure spans from $-12.3$ \kms ($\delta V=-23.1$ \kms) to $+26.7$ \kms ($\delta V=+15.9$ \kms) whilst the \ce{CH3OH} emission spans from $-3$ \kms ($\delta V=-13.8$ \kms) to $+1$ \kms ($\delta V=+1$ \kms).

First, SiO is detected towards the blue-shifted jet, as seen in previous studies (see Sect.~\ref{sec:intro}) and, for the first time, towards the red-shifted part of the jet. In our observations, the SiO jet extends up to 24$\arcsec$ ($\sim 9430$ au) to the north (blue-shifted jet) and $\sim 6\arcsec$ ($\sim$2360 au) to the south (red-shifted jet). The two jets show a relatively narrow and clumpy structure, although the northern blue-shifted SiO jet is much stronger in emission than the southern red-shifted one.  SiO emission is also seen at the systemic velocity of the source and spreads throughout the outflow cavity, from 0.4$\arcsec$ to 11.5 $\arcsec$ away from the protostar, as shown in Fig.~\ref{fig:moment-maps}.

Second, the \ce{CH3OH} ($5_{1,4}-4_{1,3}$ A; $E_{\text{u}}=49.7$ K) emission is mostly detected around the systemic velocity of the source, with emission tracing the outflow cavity walls and a final bow shock. There is no \ce{CH3OH} emission inside the outflow cavity, compared to what is seen with SiO. The \ce{CH3OH} emission centred towards the protostar tracing the hot corino \citep{bouvier2022} is also seen near the $V_{\text{sys}}$. Then, in the blue-shifted part of the outflow, a narrow ($\sim 0.4\arcsec$) \ce{CH3OH} emission is present towards the outflow cavity walls, whilst no emission is seen in the red-shifted part. 

\begin{figure*}
    \centering
    \includegraphics[width=1\textwidth]{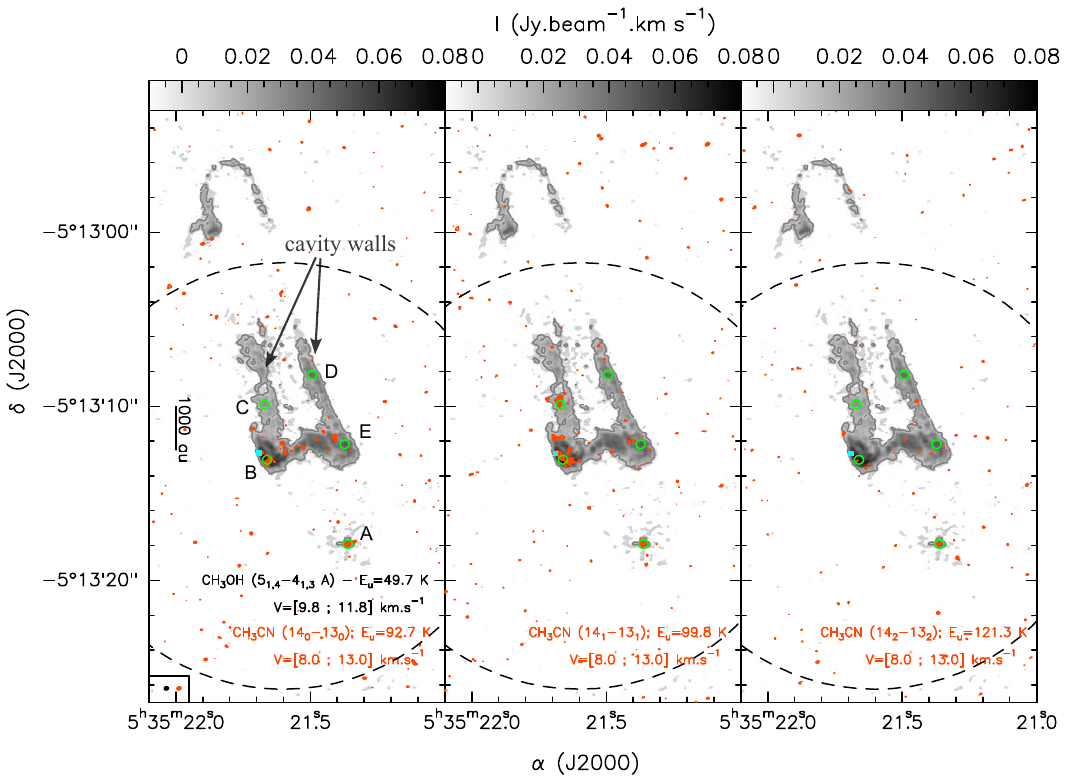}
    \caption{Moment 0 maps of \ce{CH3CN} (orange contours) overlaid on the \ce{CH3OH} ($5_{1,4}-4_{1,3}$ A) emission (grey-scaled background). The moment maps are shown only around the systemic velocity ([$10.8 \pm 1$] \kms; right), where the \ce{CH3CN} emission is detected. For \ce{CH3OH} the 4$\sigma$ level is shown with the grey contours. For \ce{CH3CN}, levels start at $4\sigma$ with steps of $1\sigma$ ($1\sigma=2.9$ mJy.beam$^{-1}$\kms). The green circles show the five regions of $0.5\arcsec\times 0.5\arcsec$ defined to extract the spectra. The filled cyan square indicates the location of the starless core FIR6c-c \citep{kainunlainen2017, bouvier2021}. The beams of \ce{CH3OH} and \ce{CH3CN} are depicted in the lower left corner of the left box. The dashed black circle represents the primary beam of 24.5$\arcsec$ from the \ce{CH3CN} observation. }
    \label{fig: CH3CN}
\end{figure*}

Both SiO and \ce{CH3OH} emission show an \textit{arc-like} feature at the base of the outflow cavity, at a distance of $\sim$ 2000 au from the source. Finally, the dust emission continuum forms a ridge almost parallel to the outflow on the east side. This ridge seems to be consistent with the dust ridge seen at a much larger scale by \citep{chini1997}, bridging between the FIR6c and FIR6a cores, which are located further to the north (outside of our field of view). Moreover, \citet{kainunlainen2017} identified a starless core within it that we previously labelled as FIR6c-c \citep{bouvier2020}. A smaller dust ridge is also seen expanding slightly to the west side of the outflow, forming a dusty \say{V}-shaped structure, likely caused by the jet. Such a feature has also been recently imaged towards the IRAS7B outflow \citep{sabatini2024}.  \\

The emission distribution of the two other methanol transitions detected in the outflow is shown in Appx~\ref{app:moment}. The emission from the \ce{CH3OH} ($1_{1,1}-1_{0,1}$ E) transition at 28 K is fainter and is concentrated to the \textit{arc-like} feature around the $V_{\text{sys}}$. The emission from this lower $E_{\text{u}}$ transition could be affected by spatial filtering. On the other hand, the emission distribution of the \ce{CH3OH} ($4_{-2,3}-3_{-1,2}$ E; $E_{\text{u}}=45.5$ K) transition resembles that of the \ce{CH3OH} ($5_{1,4}-4_{1,3}$ A; $E_{\text{u}}=49.7$ K) transition, with the emission being slightly stronger.

Lastly, the \ce{CH3CN} emission is relatively faint (in the range $3\sigma-8\sigma$) and is sporadically detected at the base of the \textit{arc-like feature} traced in SiO and \ce{CH3OH}, as well as in the outflow cavity walls seen in \ce{CH3OH}, as shown in Fig.~\ref{fig: CH3CN}. Overall, the peaks of emission of \ce{CH3CN} are consistent with those of \ce{CH3OH}. \ce{CH3CN} is only detected around the source systemic velocity. The most intense emission of \ce{CH3CN} is located towards the source position A where it could probe the hot corino, as it is the case in other low-mass protostars.\citep[e.g.,][]{bottinelli2004, taquet2015, calcutt2018a, belloche2020, nazari2021, yang2021, bianchi2022, mercimek2022, chahine2022a, hsieh2023, lee2023}.

%%%%%%%%%%%%%%%%%%%%%%%%%%%%%%%%%%%%%%%%%%%%%%%%%%%%%%%%

\section{Physical conditions and [\texorpdfstring{\ce{CH3OH}}{CH3OH}]/[\texorpdfstring{\ce{CH3CN}}{CH3CN}] abundance ratio}\label{sec:conditions}

\subsection{Derivation of the physical conditions}
We performed a non-LTE Large Velocity Gradient (LVG) analysis of \ce{CH3OH} to derive the physical conditions of the gas towards the positions B, C, D, and E shown in Fig.~\ref{fig: CH3CN}. Position A corresponds to the hot corino and the LVG analysis was performed in \cite{bouvier2022}.
The four positions selected in the outflow (B to E) are positions where the \ce{CH3OH} emission peaks. We extracted the spectra from a region of size $0.5\arcsec \times 0.5\arcsec$ (averaged spectra) and performed Gaussian fits to derive the integrated intensity ($\int T_{\text{MB}}dV$), linewidth (FWHM) and peak velocity ($V_{\text{peak}}$) of the detected lines. The spectra and Gaussian fit results are reported in Appx~\ref{app:spectra}. The results of the line fit reported in Table~\ref{tab:fits} show that the peak velocity and FWHM are larger in the outer positions C and D along the outflow with respect to positions B and E. If the outflow is jet-driven, then the velocity is expected to increase with the distance from the source \citep[e.g.,][]{raga1993, lee2001, rivera-ortiz2023}. An increase in velocity could result in an increase of turbulence and hence, in the increase of FWHM, in agreement with our findings. Alternatively, the larger line widths and different peak velocity at positions C and D could be due to the fact that part of the blue-shifted emission, which is seen close to positions C and D (Figure~\ref{fig:moment-maps}), is captured when extracting the spectra at these positions. Thus, two velocity components could be too entangled to be separated with a Gaussian fit, leading to a larger FWHM. 

To perform the non-LTE LVG analysis, we used the code \texttt{grelvg} initially developed by \cite{ceccarelli2003} and used in previous works \citep[e.g.,][]{bouvier2020, bouvier2022}. 
We used the collisional rates available in the BASECOL\footnote{\url{https://basecol.vamdc.org/}} database \citep{dubernet2013}, which were calculated by \cite{flower2010} for collision with para and ortho \ce{H2} between 10 and 200 K.  
To compute the line escape probability, we assumed a semi-infinite slab. 
We assumed a ratio \ce{CH3OH}-E/\ce{CH3OH}-A of 1, and the statistical value of 3 for the ortho-to-para ratio of \ce{H2}. 
The assumed line widths are those derived from the Gaussian fit performed on the lines (see Table~\ref{tab:fits}). 
For \ce{CH3OH}, this corresponds to $\sim 1.5$~\kms for positions B, and E, and $\sim3$~\kms and 2.5 \kms for position C and D, respectively. We included the calibration uncertainty of 10\% in the observed line intensities.

For each position, we ran a large grid of models varying the total column density of \ce{CH3OH} (which is \ce{CH3OH}-E + \ce{CH3OH}-A) in the range $10^{14}-10^{17}$ \pcms, the gas temperature between 10 and 200 K, and the gas density between $10^4$ \pcmc~ and $10^8$ \pcmc. 
These ranges were chosen to cover all the possible physical conditions we can find in a protostellar outflow. We fitted the measured line intensities simultaneously via comparison with the LVG model predictions, leaving $N_{\ce{CH3OH}}, n_{\ce{H2}}, T_{\text{kin}}$ as free parameters. For the source size, we tried both setting the size, $\theta$, to an extended emission, since the \ce{CH3OH} emission in the outflow is extended, and leaving it as a free parameter. 

\begin{figure}
    \centering
    \includegraphics[width=\linewidth]{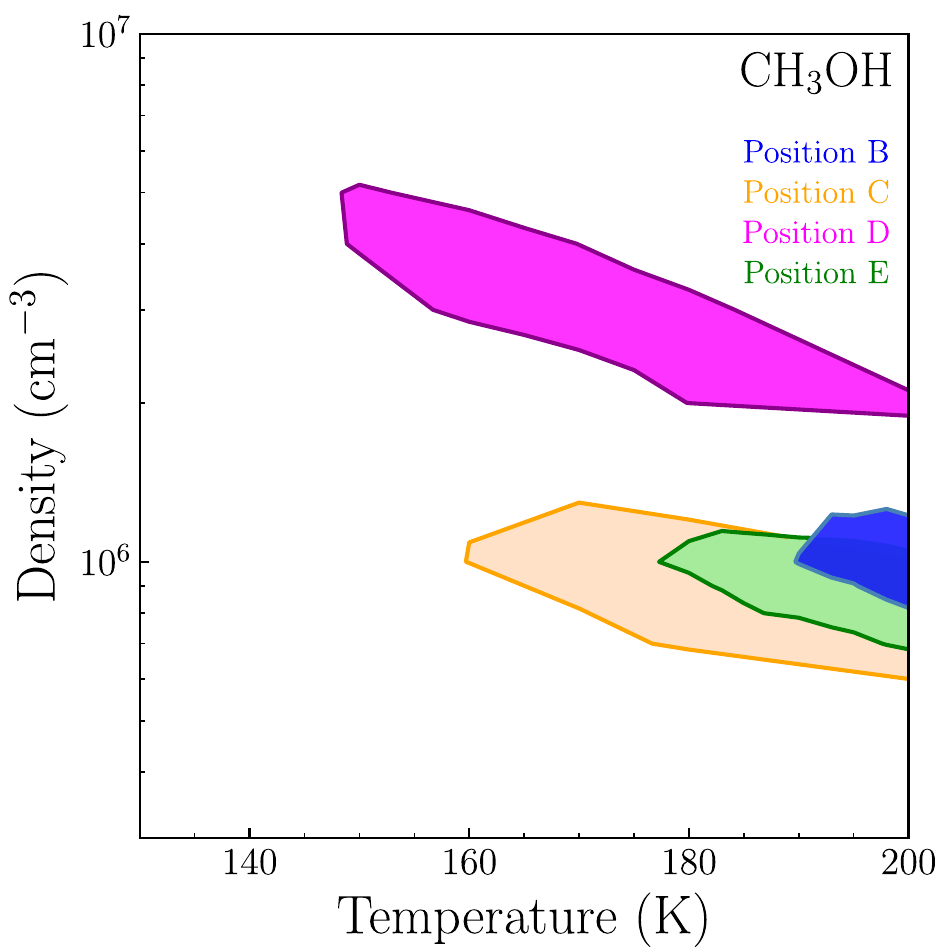}
    \caption{Density-temperature $\chi_{\text{red}}^2$ contour plots derived from the LVG analysis for positions B to E within the outflow. The contours show the 1$\sigma$ confidence level obtained for best-fit in column density and source size (see Table~\ref{tab:LVG}) for each region.}
    \label{fig:LVG}
\end{figure}

The results of the analysis are presented in Table \ref{tab:LVG}. The $1\sigma$ confidence levels constrained corresponds to a 68\% confidence level.
The best fits for the gas density and temperature at each position in the outflow are shown in Figure~\ref{fig:LVG}. 
We found that leaving the size as a free parameter leads to the best-fitted results, with lower resulting reduced $\chi^2$ ($\chi_{\text{red}}^2$). 
The best-fitted values are relatively similar in three positions (B, C and E), with the best-fit column density, gas temperature and density in the ranges of $(1.8-4)\times 10^{16}$ \pcms, $180-200$ K, and (9-10)$\times 10^5$ \pcmc, respectively.  
Position D, which is the farthest from the central object, shows denser gas with respect to the other three, with $n_{\ce{H2}}=(2-5)\times 10^6$ \pcmc. When fixing the size to an extended emission, both positions C and D show higher density compared to positions B and E, showing that the higher density found at position D compared to positions B and E, is real. The larger density estimated at the D position may be caused by the fact that the outflow encounters a local over density in the envelope at the D position, which may cause a stronger shock, hence a higher compression.
The best-fitted range of sizes of the \ce{CH3OH} emission is within the range $0.11-0.60 \arcsec$, which corresponds to a few tens of au. 
The optical depth of the line at 218 GHz is negative towards the four positions of the outflow (B to E; ($\tau=-[0.01;0.17]$), indicating that the line is a (weak) maser. 
The optical depths of the two other lines are in the range of 0.1--0.6, indicating a moderately optically thin emission. 
Overall, \ce{CH3OH} traces dense ($n_{\text{H2}}\geq 10^6$ \pcmc) and warm ($T_{\text{kin}}\geq 145$ K) gas throughout the outflow cavity.

\begin{table*}
    \centering
    \caption{Best-fit results and 1$\sigma$ confidence level (range) from the non-LTE LVG analysis, LTE results, and derived [\ce{CH3OH}]/[\ce{CH3CN}] ratio.}
    \label{tab:LVG}
    \begin{threeparttable}
    \begin{tabular}{lccccc}
    \hline \hline
& Position A\tnote{a} & Position B & Position C & Position D & Position E \\
\hline
\multicolumn{6}{c}{LVG results: \ce{CH3OH}}\\
\hline
$n_{\text{H2}}$ [$\times 10^5$ \pcmc]  best-fit  &40 & 10 & 10 &30 &9  \\
$n_{\text{H2}}$ [$\times 10^5$ \pcmc]   range & $30-50$ &$8-15$ &$6-15$&$20-50$&$6-15$ \\
$T_{\text{kin}}$ [K] best-fit &180 & 198 & 180 &170 &200\\
$T_{\text{kin}}$ [K] range & $\geq 85$ & $\geq 190$ &$\geq 160$ &$\geq 145$&$\geq 175$ \\
$N_{\text{tot}}$ [$\times 10^{16}$ \pcms] best-fit & 120& 1.8 & 4 &1.4 &1.9 \\
$N_{\text{tot}}$ [$\times 10^{16}$ \pcms] range & $80-200$ & $1.4-3$ &$1.2-8$ &$0.6-3$&$1.4-3$  \\
Size [$\arcsec$] best-fit & 0.10 & 0.24 & 0.15 &0.28& 0.18\\
Size [$\arcsec$] range & $0.07-0.13$ & $0.17-0.30$&$0.11-0.30$&$0.15-0.60$&$0.14-0.24$\\
\hline
\multicolumn{6}{c}{LTE results: \ce{CH3CN}}\\
\hline
Size [$\arcsec$] range used\tnote{b} &  $0.07-0.13$ & $0.17-0.30$&$0.11-0.30$&$0.15-0.60$&$0.14-0.24$\\
T [K] used\tnote{b} &180 & 198 & 180 &170& 200 \\
$N_{\text{tot}}$ [$\times 10^{14}$ \pcms]& $4.4-15.8$ & $0.9 -1.1$ & $0.8-4.9$ &$\leq$ 2.9 &$0.5-1.4$ \\
\hline
[\ce{CH3OH}]/[\ce{CH3CN}] &$506-4545$ & $127-326$ & $24-1000$ &$\geq$ 20&$93-600$ \\
\hline
    \end{tabular}
    \begin{tablenotes}
    \item[a]The LVG results are from \cite{bouvier2022}
    \item[b]Using the \ce{CH3OH} LVG results
    \end{tablenotes}
    \end{threeparttable}
\end{table*}

\subsection{[\texorpdfstring{\ce{CH3OH}}{CH3OH}]/[\texorpdfstring{\ce{CH3CN}}{CH3CN}] abundance ratio}

For \ce{CH3CN}, although we detect the K=0-2 ladders towards positions A and E, the range of $E_{\text{u}}$ covered is too narrow to perform the LVG analysis. 
At positions B and C, we detect the K=0-1 and K=1 ladders, respectively. The spectra are shown in Appx~\ref{app:spectra}. Position E is where the detection of \ce{CH3CN} is the faintest  
No \ce{CH3CN} is detected at position D. 
To compute the total column density of \ce{CH3CN}, $N_{\ce{CH3CN}}$, we assume that in all positions, both \ce{CH3OH} and \ce{CH3CN} are emitted from the same region. Therefore, we used the rotational diagram method with a fixed size range and a fixed temperature, taken from the LVG result of \ce{CH3OH}, to derive $N_{\ce{CH3CN}}$. As the gas temperature derived for \ce{CH3OH} are lower limits, we used the best-fit temperature of each position. As a result, we obtained $N_{\ce{CH3CN}}=(4.4-15.8)\times 10^{14}$ \pcms \ towards the hot corino and a range in $N_{\ce{CH3CN}}$ of $(0.5-4.9)\times 10^{14}$ \pcms \ in the outflow. We then derived the abundance ratio [\ce{CH3OH}]/[\ce{CH3CN}] towards the hot corino (position A) and the three positions in the outflow (positions B, C, E) and found an abundance ratio of $506-4545$, $127-326$, $24-1000$, and $93-600$ towards the hot corino, the position B, C, and E, respectively. For Position D, as \ce{CH3CN} is not detected, we calculated the upper limit of the column density of \ce{CH3CN}. To do so, we calculated the $3\sigma$ upper-limit level of the integrated intensity (in K.\kms) using the formula $3\times\text{rms}\times\text{FHWM} \times\sqrt{N_{\text{chan}}}$, where $N_{\text{chan}}$ is the number of channel within the linewidth ($\text{FWHM}$). We assumed a linewidth of 2.0~\kms, mean value found for \ce{CH3CN} at Position C. We then used the derived upper limit column density to calculate a lower limit for the abundance ratio [\ce{CH3OH}]/[\ce{CH3CN}]. The results are summarised in Table~\ref{tab:LVG}.

%-----------------------------------------------------------

\section{Characterisation of the SiO jets}\label{sec:jet}

Thanks to the high angular resolution achieved with the ORANGES observations, we saw in Sec.~\ref{sec:emission-distribution} that we can resolve the knotty structure of the SiO jets. This allows us to investigate the collimation angle of the two jets, and quantify the episodic timescale of the ejections.

\subsection{Jet collimation angle}\label{subsec: collimation}

\begin{figure*}
     \centering
    \includegraphics[width=0.85\textwidth]{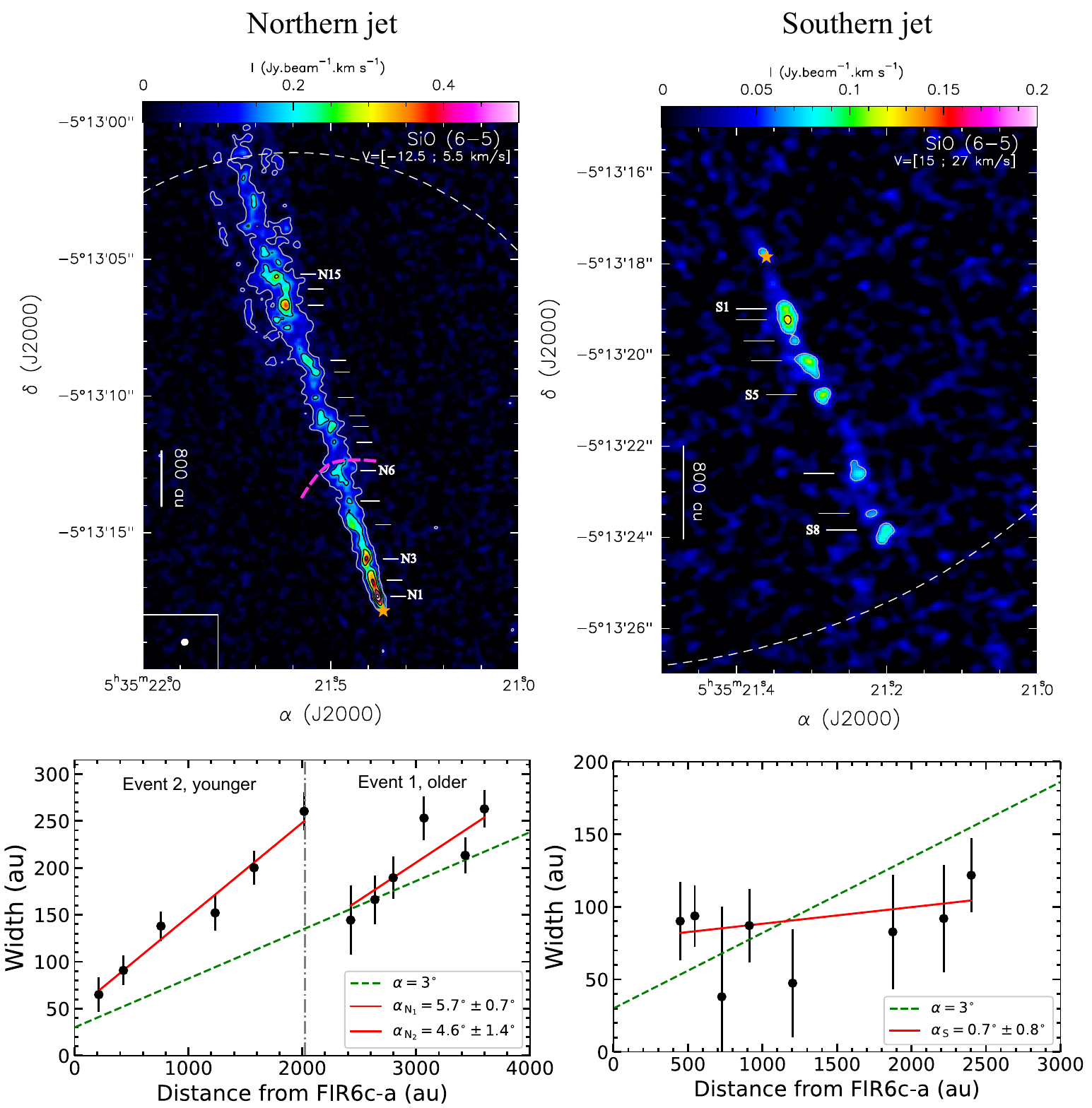}
    \caption{\textit{Top:} Velocity integrated emission of the SiO (6-5) emission associated with the jet zoomed-in towards each jet. 
    The position of the knots along the jet are indicated by the white horizontal lines. 
    The position of the protostar FIR6c-a is marked by the filled orange star. 
    The primary beam is shown by the white dashed line and the beam is depicted in the bottom left part of each box. 
    \textit{Left upper panel:} Emission from the northern (blue-shifted) part of the jet, integrated between -12.5 and +5.5 \kms. Contours start at  10$\sigma$ ($1\sigma=$7 mJy.beam.\kms) and increase with steps of 15$\sigma$.
    \textit{Right upper panel:} Emission from the southern (red-shifted) part of the jet, integrated between +15 and +27 \kms. Contours start at 10$\sigma$ ($1\sigma=$5.7 mJy.beam.\kms) and increase with steps of 20$\sigma$.\
    \textit{Bottom panels:} Deconvolved widths ($2R_{\text{jet}}$) of the SiO (6-5) emission as a function to the distance from FIR6c-a, for the northern jet (\textit{left}) and the southern jet (\textit{right}). 
    The solid red lines correspond to the fit performed on the data points (black filled circles). The dashed green lines have an opening angle of $3^{\circ}$ and initial width of 30 au,  corresponding to the typical collimation found in Class II atomic jets \citep[e.g.,][]{dougados2000, woitas2002, agraamboage2011}. 
    The position of the \textit{arc-like} feature (see Sec.~\ref{sec:emission-distribution}) is indicated by a dashed magenta line in the top left panel and by a grey dashed and dotted line in the bottom left panel.} 
    \label{fig:collimation-angle}
\end{figure*}

Figure~\ref{fig:collimation-angle} shows a zoom on the two SiO jets of the system, north (blue-shifted with respect to the systemic velocity) and south (red-shifted).
To investigate the collimation angle of the two SiO jets we followed the methodology presented in \citet{chahine2022b}. Contours in Figure~\ref{fig:collimation-angle} show where the SiO emission is $> 10\sigma$, which we used to identify the knots along the two jets. We identified fifteen and eight knots towards the blue- and red-shifted jets, respectively. 
Then, we performed perpendicular cuts, and we measured the observed widths, $\text{FWHM}_{\text{obs}}$ as described in \citet{chahine2022b}.
We then calculated the deconvolved widths, $2R_{\text{jet}}$, using the equation from \citet{podio2021}:

\begin{equation}
        2R_{\text{jet}}=\sqrt{\text{FWHM}^2_{\text{obs}}-b_t^2}
\end{equation}

where $b_t$ is the transverse size of the beam (see Eq. D.2 from \citealt{podio2021}).
For the northern blue-shifted jet we selected only the knots 1 (the closest to the protostar) to 12, as we could get a clear spatial profile for these knots only.  
Finally, we calculated the collimation angle of the two jets, by fitting the following equation:

\begin{equation}
    2R_{\text{jet}}=2tan(\alpha/2)z+2R_0
\end{equation}
where $z$ is the distance from the source FIR6c-a, $\alpha$ is the apparent opening of the jet, and $R_0$ is a constant offset. 

As in \citet{chahine2022b}, we plotted the derived deconvolved width as a function of the distance from the central object FIR6c-a (Fig. \ref{fig:collimation-angle}) and drew a reference line, representing a collimation angle of $\alpha=3^\circ$ and $R_0$=30 au \citep{podio2021}. 
Interestingly, for the northern jet, there is a discontinuity in the calculated width occurring at $\sim 2000$ au from the protostar. 
The location of the discontinuity occurs at the location of the \textit{arc-like} feature identified in Sec.~\ref{sec:emission-distribution} in the SiO and \ce{CH3OH} emissions. 
Since jets are known to be episodic events \citep[e.g.,][]{arce2007, frank2014}, we hypothesise that the northern jet is actually due to two distinct events, an older one, Event 1, and a younger one, Event 2, where the \textit{arc-like feature} marks the end of the youngest one (Event 2). 
For both events, the width clearly increases with the distance, implying an opening up of the jets.
We derived the two collimation angles, corresponding to the two different events, to be $(5.7\pm0.7)^\circ$ and $(4.6\pm 1.4)^\circ$ for the event 2 and 1, respectively. 
Hence, within the error bars, the two jets have the same collimation angle. 
Since such a feature is not seen in the southern jet, we calculated only one collimation angle corresponding to a single event. 
For the southern jet, the widths increase only slightly with the distance and we derive a full opening angle of $(0.7 \pm 0.8) ^\circ$. 
Even with the relatively high error, we can put an upper limit to the collimation angle of $<1.5^\circ$, which is smaller than the two in the northern jets. We discuss the possible causes for having a narrower opening angle in the southern jet in Sec.~\ref{subsec:dichotomy}.

\subsection{Dynamical timescale of the knots}\label{subsec:tdyn}

We calculated the dynamical timescale of the knots N1, N3, N6, N15 and S1, S5, and S8 of the northern and southern jets, respectively. A few bright knots were selected to give an overview of the typical dynamical timescales along the northern and southern jets. The knots for which the dynamical age is derived are indicated in Fig.~\ref{fig:collimation-angle}. 
We used the following formula:
\begin{equation}
    \tau_{dyn}=\frac{d_{knot}}{V_{jet}}
\end{equation}
where $d_{knot}$ is the distance of the knot from the driving source, $V_{jet}$ is the typical jet velocity. We assume that the typical velocity of the jet is 100 \kms \citep[see][and refs therein]{podio2021}. The deprojected distance of the knot, $d_{knot}$ is $d_{knot}=d_{pos}/sin (i)$, with $d_{pos}$ being the knot distance measured on the plane of the sky and $i$ the inclination angle of the outflow with respect to the line of sight. From the literature, the inclination of the outflow is thought to be $\sim 50^\circ$ \citep{tobin2016b}.
Therefore, in the northern jet, the distances of the knots N1, N3, N6 and N15 on the plane of the sky are $d_{N_1}\sim 210$ au, $d_{N_3}\sim 760$ au, $d_{N_6}\sim 2000$ au, $d_{N_{15}}\sim 4800$ au, and the corresponding dynamical ages are $\tau_{dyn,N1}=$13 yr, $\tau_{dyn,N3}=$41 yr, $\tau_{dyn,N6}=$124 yr, and $\tau_{dyn,N15}=$298 yr, respectively. 
In the southern jet, the distances of knots S1, S5, and S8 on the plane of the sky are $d_{S_1}\sim450$ au, $d_{S_5}\sim1200$ au, and $d_{S_8}\sim2400$ au with dynamical ages of $\tau_{dyn,S1}=$28 yr, $\tau_{dyn,S5}=$74.5 yr, and $\tau_{dyn,S8}=$149 yr, respectively.

\begin{figure}
    \centering
    \includegraphics[width=0.4\textwidth]{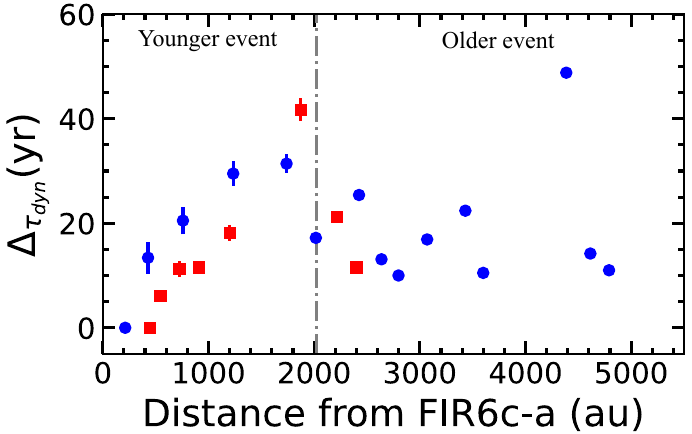}
    \caption{Dynamical time interval ($\Delta\tau_{dyn}$) between subsequent knots as a function of the distance from FIR6c-a, for the northern blue-shifted jet (filled blue circles), and southern red-shifted jet (filled red squares). The location of the \textit{arc-like} feature is indicated by the dashed and dotted grey line at $\sim$ 2000 au. We indicated the different possible events of jet ejection, from the younger (Event 1) to the older (Event 2).}
    \label{fig:jet-tdyn-interv}
\end{figure}

Figure~\ref{fig:jet-tdyn-interv} shows the timescale interval between the blue- and red-shifted knots as a function of the distance to FIR6c-a. 
In the blue-shifted side of the jet, we see an increasing $\Delta\tau_{dyn}$  up to 2000 au where the youngest (Event 2) jet ends (see Sec.~\ref{subsec: collimation}). 
Beyond 2000 au, no clear trend is seen in the timescale of the ejections. 
In the red-shifted side of the jet, $\Delta\tau_{dyn}$ also increases up to $\sim$ 2000 au before decreasing for the last two knots. 
If the north and south jet events are symmetric, then it could indicate that the last two knots are not associated with the same ejection event as the other knots and that also the southern jet, after all, is due to the two events.
Finally, the dynamical time interval between the knots, on both the northern and southern jets, shows a relatively short period of time for the ejections, of a few tens of years. 
Such short periods have been suggested to happen in protostellar jets \citep[e.g.,][]{IF2012}.

%------------------------------------------------------------

\section{Astrochemical modelling}\label{sec:model}

To derive some information about the age of the outflow, we tried to use the chemical behaviour of \ce{CH3OH} and \ce{CH3CN}, specifically, the
measured [\ce{CH3OH}]/[\ce{CH3CN}] abundance ratio compared to theoretical model predictions.
To do this, we leveraged the study by \citep{giani2023}, who carried out a very extensive review of the gas-phase reactions involved in the formation of CH$_3$CN and on the assumption that, on the contrary, methanol is mostly formed on the grain surfaces and released in the gas-phase by the shock passage. 

\begin{table}
\renewcommand{\arraystretch}{1.3}
\centering
    \caption{The left half table lists the initial elemental abundances relative to H nuclei adopted for the cold molecular cloud modelling \citep{jenkins2009unified}. The right half table lists the abundance of the species injected in the gas-phase after the shock passage. The adopted values are the same as those derived in a similar molecular shock, L1157-B1, based on astronomical observations \citep{podio2014, codella2017, codella2020, giani2023}. }
    \label{tab:model-initial+injected}
    \begin{tabular}{cc|cc}
    \hline
    \hline
    \multicolumn{2}{c|}{Initial abundances} &  \multicolumn{2}{c}{Injected abundances} \\
    \hline
    Element  & Abundance &  Species & Abundance\\
    \hline
    He      & $9.0 \times 10^{-2}$  & H$_{2}$O    & $1\times 10^{-4}$ \\
    C$^+$   & $1.7 \times 10^{-5}$  & CO$_{2}$    & $3\times 10^{-5}$ \\
    O       & $2.6 \times 10^{-5}$  & CO    & $1\times 10^{-4}$ \\
    N       & $6.2 \times 10^{-6}$  & CH$_3$OH  & $0.1-10\times 10^{-6}$ \\
    S$^+$   & $8.0 \times 10^{-8}$  & NH$_3$    & $5.6\times 10^{-5}$ \\
    Si$^+$  & $8.0 \times 10^{-9}$  & H$_{2}$CO   & $1\times 10^{-6}$ \\
    P$^+$   & $2.0 \times 10^{-10}$ & OCS       & $2\times 10^{-6}$ \\
    Na$^+$  & $2.0 \times 10^{-9}$  & SiO       & $1\times 10^{-6}$ \\
    Mg$^+$  & $7.0 \times 10^{-9}$  & Si        & $1\times 10^{-6}$\\
    Fe$^+$  & $3.0 \times 10^{-9}$  & CH$_3$CH$_{2}$ & $4\times 10^{-8}$  \\
    Cl$^+$  & $1.0 \times 10^{-9}$  & CH$_3$CH$_{2}$OH & $6\times 10^{-8}$ \\
    F$^+$   & $1.0 \times 10^{-9}$  & SiH$_4$  & $1\times 10^{-7}$ \\
    \hline
    \end{tabular}
\end{table}

\subsection{Model description and adopted parameters}
To compute the theoretical abundances of \ce{CH3OH} and \ce{CH3CN}, we ran the MyNahoon code, an in-house modified version of the publicly available Nahoon code \citep{wakelam2005estimation,wakelam2010sensitivity}. 
Briefly, the code computes the gas-phase abundances for given physical parameters (T, $n_{\ce{H2}}$, $A_v$ and $\zeta_{CR}$) using the GRETOBAPE gas-phase network \citep{tinacci2023-gretobape}. 
Surface reactions are not taken into account, except for the formation of \ce{H2}.

To simulate the impact on the chemical composition due to the passage of a shock, we proceeded in two steps, following a method already used by our group to model the protostellar molecular shock L1157-B1 \citep[e.g.,][]{codella2017, codella2020, giani2023}. 
As a first step, we compute the gas composition of the cloud before the passage of the shock, starting from the elemental abundances reported in Tab. \ref{tab:model-initial+injected} and assuming T=10 K, $n_{\ce{H2}}=1\times10^4$ \pcmc, $A_v=100$ mag and $\zeta_{CR}=1\times 10^{-17}$ \s. 
In the second step, we increased the abundance of several species that are sputtered from the dust grain mantles (see Tab. \ref{tab:model-initial+injected} for the list of the injected species and their abundances) and we compute again the evolution of the gas composition for a gas temperature and density of 180 K and $8\times10^5$ \pcmc, respectively. 
The injected methanol abundance is varied between $1\times10^{-7}$ and $1\times10^{-5}$, to see the effect on the methyl cyanide formation. 
Moreover, we used three different values of CR ionisation rate ($\zeta_{CR}=1\times 10^{-16},1\times 10^{-15},1\times 10^{-14}$ \s).
Given the similar physical conditions of the three positions of the outflow B, C, and E, we consider this model as representative of all these three regions. 
Finally, we computed the evolution of the [\ce{CH3OH}]/[\ce{CH3CN}] ratio as a function of the injected methanol abundance for the three $\zeta_{CR}$ values and three different times after the shock passage (100, 500 and 1000 yr, respectively).

\subsection{Results of the modelling}

\begin{figure}
    \includegraphics[width=0.5\textwidth]{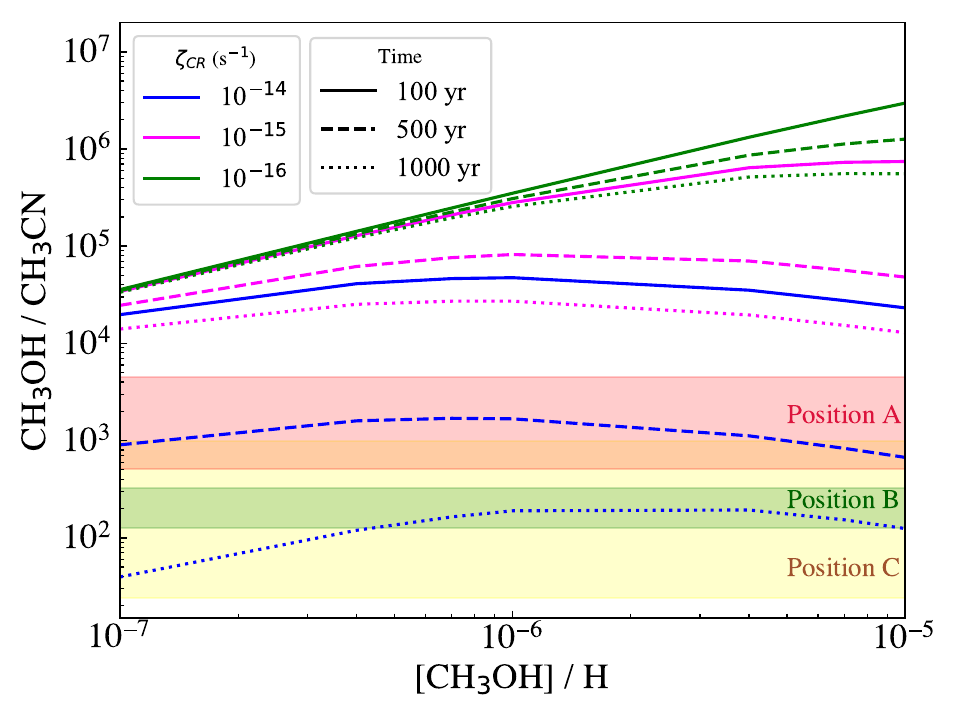}
    \caption{Comparison of predicted and observed [\ce{CH3OH}]/[\ce{CH3CN}] abundance ratio as a function of the CH$_3$OH abundance (with respect to H nuclei). 
    The solid, dashed and dotted lines correspond to model predictions at different times after the shock passage: 100 (solid line), 500 (dashed line) and 1000 (dotted line) yr, respectively. 
    The predicted [\ce{CH3OH}]/[\ce{CH3CN}] ratio is also reported for three different values of $\zeta_{CR}$: $1\times10^{-14}$ (blue lines), $1\times10^{-15}$ (magenta lines) and $1\times10^{-16}$ (green lines) \s.
    The shaded bands show the observed [\ce{CH3OH}]/[\ce{CH3CN}] abundance ratio (Tab. \ref{tab:LVG}), for the three positions A (red), B (green) and C (yellow). The values for position E are not represented but are similar to those for positions B and C, as shown in Table~\ref{tab:LVG}.}
    \label{fig:model-CH3OH-CH3CN}
\end{figure}

Figure \ref{fig:model-CH3OH-CH3CN} shows the time-dependent [\ce{CH3OH}]/[\ce{CH3CN}] abundance ratio as a function of the \ce{CH3OH} injected abundance, as predicted by the model for different values of $\zeta_{CR}$. 
The model predictions are also compared to the LVG results reported in Tab. \ref{tab:LVG}. 
Note that, since methanol is destroyed once injected into the gas phase, its injected abundance does not necessarily correspond to the abundance observed in the gas phase. 
Thus, the observed \ce{CH3OH} abundance should be considered as a lower limit to the injected one. 
However, because of the lack of an estimate of the \ce{H2} column density, it was not possible to directly compare the injected and observed methanol abundances.

The [\ce{CH3OH}]/[\ce{CH3CN}] ratio also depends on the injected methanol abundance, as for low \ce{CH3OH} abundances the \ce{CH3+} ion, which is the bottleneck for the formation of \ce{CH3CN}, is formed by the \ce{CH2+} + \ce{H2} reaction, while for large \ce{CH3OH} abundances the \ce{CH3+} ion is formed by the reaction of methanol with \ce{H3+} \citep{giani2023}. 
We do not expect significant variations of the modelling predictions for different injected abundances of the other species.
With increasing time, the [\ce{CH3OH}]/[\ce{CH3CN}] abundance ratio decreases as the injected methanol is destroyed by reactions with molecular ions, such as \ce{H3+} or \ce{H3O+}. 
At the same time, \ce{CH3CN} abundance increases with time, as it is efficiently formed because of the larger availability of the reactant species in the gas phase. 
Specifically, \ce{CH3CN} is formed in a two-step process: the radiative association of \ce{CH3+} + HCN  to form \ce{CH3CNH+}, followed by the recombination with an electron or the proton transfer to ammonia. 

Increasing the value of $\zeta_{CR}$ has the effect to decrease the [\ce{CH3OH}]/[\ce{CH3CN}] abundance ratio.
For $\zeta_{CR}=1\times10^{-16}$ \s, methanol is mainly destroyed by the reaction with \ce{H3+}, providing large quantities of \ce{CH3+} ion which can be used as the reactant to form \ce{CH3CN}. 
When increasing $\zeta_{CR}$ up to $1\times10^{-14}$ \s, methanol is more efficiently destroyed by \ce{H3O+} and the reaction \ce{CH2+} + \ce{H2} becomes the main source of \ce{CH3+}. 
The enhanced destruction of methanol combined with the increased formation of methyl cyanide leads to a decrease of the [\ce{CH3OH}]/[\ce{CH3CN}] abundance ratio by several orders of magnitude.

When comparing the measured [\ce{CH3OH}]/[\ce{CH3CN}] abundance ratio with the model, we find two main results. 
First, the abundance ratios derived in the outflow, hence in positions B and C, are reproduced by the model at a time equal to 1000 yr for most of the range of the methanol abundance. 
The ratio at position C could be consistent with an age of 500 yr for [\ce{CH3OH}/H] $\geq 4\times10^{-6}$ or $\leq 2\times10^{-7}$. 
Second, the abundance ratios measured in the outflow and the hot corino are all reproduced only for a cosmic-ray ionization rate of $\zeta_{CR}=10^{-14}$ \s, hence 1000 times more than the average galactic value.

%------------------------------------------------------------

\section{Discussion} \label{sec:disc}
\subsection{A new iCOM-rich outflow}\label{subsec:chemicalrichness}

One of the main interesting features of the outflow driven by the FIR6c-a outflow is the detection of \ce{CH3OH} and \ce{CH3CN}, which makes the FIR6c-a outflow the second iCOM-rich outflow detected in the OMC-2/3 filament, after HOPS-87 where several species including \ce{CH3OH} were recently detected at the base of its molecular outflow \citep{hsu2024}. The FIR6c-a outflow thus adds itself to the relatively poor number of iCOM-rich known outflows. Indeed, these two iCOMs, as well as a few other ones, such as acetaldehyde (\ce{CH3CHO}), formamide (\ce{NH2CHO}), dimethyl ether (\ce{CH3OCH3}), methyl formate (\ce{HCOOCH3}), and ethanol (\ce{C2H5OH}), have been detected in a low-mass protostellar outflows and associated shocks (e.g.,\citealt{sandell1994, bachiller1995, bachiller1998, gibb1998, arce2008, tafalla2010, oberg2011, codella2017, codella2020, holdship2019b}, Robuschi et al. in prep.).

Towards the FIR6c-a outflow, \ce{CH3OH} is detected towards the blue-shifted cavity walls as well as in the two bow shocks (see discussion in Sec.~\ref{subsec: arc-like feature}). 
The physical parameters derived in Sec.~\ref{sec:conditions} show that the methanol lines trace relatively dense ($n_{\text{H2}}\geq 5\times10^5$ \pcmc) and hot gas ($T_{\text{kin}}\geq 145$ K) in positions B to E. At these positions, we found that the \ce{CH3OH} ($4_{-2,3}-3_{-1,2}$E) transition is a weak maser (from the negative optical depth derived in Sec.~\ref{sec:conditions}). 
Class I-type methanol masers are known to be associated with protostellar outflows and are proposed to be collisionally excited, as in our case, because of the enhanced density due to the interaction between the outflow and the surrounding medium \citep[e.g.,][]{PM90, johnston1992, sandell2005, voronkov2006, kalenskii2010, gan2013, ladeyschikov2020}.  
Therefore, methanol very likely traces shock spots at positions B to E within the FIR6c-a outflow. 
Moreover, this particular transition is already known as a class I-type maser \citep{hunter2014} and several maser spots were previously found in the FIR4 region, north to FIR6c-a \citep{chahine2022b}, although the maser spots were found in filaments rather than outflows in this region. 
In conclusion, we report the presence of several shock spots with class-I type methanol maser in the FIR6c-a outflow. 

Concerning \ce{CH3CN}, its emission is quite faint and peaks mostly where there are \ce{CH3OH} peaks, i.e., in the first bow-shock represented by the \textit{arc-like} feature and along the cavity walls, indicating that \ce{CH3CN} traces the shock spots as \ce{CH3OH}.  
This is similar to what was found towards L1157-B1, where \cite{codella2009} concluded that \ce{CH3CN} traces warm ($T_{\text{kin}}=60-130$ K) shocked gas. 
\ce{CH3CN} was also found to probe shocked gas towards other star-forming regions \citep[e.g.][]{csengeri2011, leurini2011, bell2014}. 

Finally, towards the FIR6c-a outflow (positions B to E), we derive a [\ce{CH3OH}]/[\ce{CH3CN}] ratio in the range 24$-$1000, consistent with that derived towards L1157-B1 (5$-$769; \citealt{codella2009}), and towards IRAS 20126+4104 ($\sim 189$; see Table 1 of \citealt{palau2017}). 

Comparing our results with those from the L1157-B1 and -B2 shock positions, where the physical conditions of the gas tracing \ce{CH3OH} have been derived via non-LTE modelling \citep{mcguire2015,holdship2019b, codella2020}, we find that our results are similar for the density but the temperature in FIR6c-a is higher than that found in L1157-B1 and -B2 shocks (40--130 K). On the other hand, other tracers towards L1157-B1 probe higher temperatures, similar to the one we derived in FIR6c-a \citep[e.g.,][]{benedettini2012, lefloch2012, feng2020}. In the same vein, \ce{CH3OH} was found to trace gas with temperatures up to 155 K and densities up to $5\times10^7$\pcmc towards the outflow of HOPS 373SW \citep{lee2024} and a gas temperature of  40--130 K in a sample of seven low-mass protostellar outflows
\citep{holdship2019b}, hence similar to that derived in L1157-B1 and colder than that in the FIR6c-a outflow. 
In this latter case, however, the temperature difference could be attributed to the fact that \cite{holdship2019b} used single-dish observations (i.e., probed a larger area possibly contaminated by the colder gas in the envelopes of the targeted sources). On the other hand, the temperature we derive here is similar to that derived in the outflow of the high-mass protostar IRAS 20126+4104 ($\sim 200$ K; \citealt{palau2017}). To understand what is the main source of gas heating along the outflow, we can assess the gas temperature that could be heated by the source directly. At the densities derived in this work ($n_{\mathrm{\ce{H2}}}> 10^5$ \pcmc), we can consider that the dust and the gas are coupled. Hence, using Equation 1 from \cite{ceccarelli2000a}, for a source luminosity of 8.2 L$_\odot$ at 150 au, the gas temperature is $\sim 56$ K. At positions B to E, which are located $\geq 2000$ au from the source and where we derived $T_{\mathrm{gas}}>140$ K, the main source of gas heating along the outflow is shocks rather than the source's luminosity.

Finally, the detection of \ce{CH3OH} and \ce{CH3CN} within the FIR6c-a outflow makes this outflow a new astrochemical laboratory to study chemical richness in protostellar outflows. A future study will be dedicated to fully characterising the chemical richness of the FIR6c-a outflow.

\subsection{A clear dichotomy between the northern and southern outflow}\label{subsec:dichotomy}

The blue-shifted and red-shifted outflows driven by FIR6c-a show a clear dichotomy in molecular emissions, already noted by previous studies. 
First, \cite{shimajiri2009} found the presence of SiO jets only towards the northern blue-shifted outflow and concluded that the non-detection of the red-shifted SiO jet was due to insufficient dense gas material in the south and, hence, the amount of dust grains available is not sufficient to produce detectable SiO column densities \citep{bachiller2001}. 
Later, \cite{tobin2016b} observed CO (1-0) with CARMA and \ce{H2O} with the \textit{Herschel Space Observatory}, and found differences between the blue-shifted and red-shifted outflows: whilst the blue-shifted outflow coincides with the region emitting high-J ($J_{u}> 13$) CO and \ce{H2O} lines, the red-shifted outflow does not show emission from high-J CO or \ce{H2O}. 
The authors also concluded that the difference was likely because whilst the northern blue-shifted outflow propagates into a dense ambient medium causing shocks, the medium in the southern region is less dense. 
Both \cite{shimajiri2009} and \cite{tobin2016b} supported their hypothesis with the fact that extended cold dust emission is observed towards the north of the outflow but not the south.

In our observations, for the first time, we detect also the red-shifted part of the SiO jet. 
However, we also see a clear difference between the blue- and red-shifted jet with the red-shifted jet less intense and reaching only $\sim 6\arcsec$ ($\sim 2300$ au) away from the protostar whilst the northern jet propagates up to $\sim 24\arcsec$ ($\sim 9400$ au; although it is known to extend beyond our field of view; e.g., \citealt{stanke2002, takahashi2008, shimajiri2009, tobin2016b, gomezruiz2019}). 
In addition to the SiO jet's length, the derived collimation angle differs in the northern and southern jets: whilst the two blue-shifted SiO jets (see Sec.~\ref{subsec: arc-like feature}) have a collimation angle close to $5^\circ$ on average, the red-shifted SiO jet has a collimation angle $\leq 1.5^\circ$. 
Asymmetric or mono-polar jets have been observed towards several other sources \citep[e.g.,][]{melnikov2009, codella2014a, fernandez-lopez2013, nony2020, podio2021, chahine2022b, dutta2024} although the cause is not clear. 
A first possibility, already invoked earlier, is an asymmetric ambient medium in which the two sides of the outflow propagates: a fainter or a lack of SiO jet could be explained by the fact that it propagates in a cavity already \textit{cleaned-out} by previous ejection events, hence in a lower density medium, or by its deflection after hitting a dense clump of gas and dust \citep[e.g.,][]{raga2002, melnikov2009, fernandez-lopez2013} which can be reasonable in a high-mass star-forming region such as Orion. An asymmetric ambient medium can also lead to an asymmetric interaction of the disc with its surroundings, thus leading to a difference in the region of the disk launching the jet \citep[e.g.,][]{melnikov2009, ferreira2006}. In our case, the hypothesis of a less dense medium could be viable considering that we detect cold dust emission towards the north of the outflow (see Fig.~\ref{fig:moment-maps}) but not in the south, as seen already in previous studies at larger scales \citep{shimajiri2009, tobin2016b}. If the red-shifted jet evolves in a less dense medium, it could propagate faster as it interacts less with the ambient medium, thus leading to a higher collimation of the jet \citep{podio2011}, and its faintness can thus be explained by a smaller dust column density in the region.

Another explanation for asymmetric or mono-polar jets is directly linked to their launching mechanism, where the ejection of material from the disk is not homogeneous \citep[e.g.][]{matsakos2012, FS13, codella2014a, bai2017, bethune2017, zhao2018}. However, despite the differences in the lengths and collimation angles, the two jets seem to have been created by events that symmetrically ejected material towards the north and south, as shown by Figure~\ref{fig:jet-tdyn-interv}, both in the younger ( at a distance $\leq$2000 au) and older ejections. Indeed, the two jets show the same \say{break-down} of the dynamical time interval at $\sim 2000$ au. We could thus rule out the hypothesis for an asymmetric launching mechanism. Lastly, another possible explanation of the smaller collimation angle of the southern jet with respect to the northern one is just a bias in the observations, and only a fraction of the jet is actually detected due to the smaller column density of the shocked material. Observations with higher sensitivity could help understand if fainter emission is missed.

Finally, another clear differentiation between the northern and southern outflow is the clear chemical difference of the two cavities: 
Whilst both \ce{CH3OH} and \ce{CH3CN} are detected in the northern cavity, none of these species are detected in the southern one. 
Additionally, the SiO is not only detected in the jet but also in the blue-shifted outflow cavity, surprisingly. 
We will address the origin of the chemical differentiation of the two outflow cavities in a future study. 

\subsection{Footprints of intricate temporal outflowing events and origin of the arc-like feature}\label{subsec: arc-like feature}

Since we detected both \ce{CH3OH} and \ce{CH3CN} towards several shock spots within the FIR6c-a outflow (see Sec.~\ref{subsec:chemicalrichness}), we tried to constrain the shock age using the [\ce{CH3OH}]/[\ce{CH3CN}] abundance ratio as done in \cite{giani2023} for the L1157-B1 shock. 
In the latter case, \cite{giani2023} found that the observed ratio could be reproduced for shock ages larger than 1000 yr, consistent with what previous studies derived \citep{gueth1996, podio2016}, so the method seems to be quite reliable. In addition, using iCOMs emission towards the high-mass protostellar outflow of IRAS 20126+4104 and coupling a shock model with a chemical network, \cite{palau2017} could also constrain the age of the outflow to about 2000 yr.
In the case of FIR6c-a, the predicted [\ce{CH3OH}]/[\ce{CH3CN}] abundance ratio point to timescales $\geq$1000 yr for the shocks in positions B, C, and E (see Sec.~\ref{sec:model}). 
This is relatively consistent with previous estimations of the outflow dynamical age by \cite{takahashi2008} and \cite{tobin2016b}, who estimated an age of $\sim (1.2-1.9)\times 10^4$ yr and $\sim 5000$ yr, respectively.

Outflow and jet ejections are known to be episodic events \citep[e.g.,][]{frank2014, lee2020}. 
The formation of the knots is still debated and could be due to a periodic variation in the jet velocity or a variation in the episodic mass ejections \citep[e.g.][]{plunkett2015, JL16, matsushita2019,jhan2022, dutta2024}. 
The dynamical timescale between the knots varies from a few years to a few hundred years \citep[e.g.,][]{plunkett2015, lee2020, jhan2022}. 
In the FIR6c-a case, the period of variation that we derived (Fig.~\ref{fig:jet-tdyn-interv}) suggests that subsequent ejections occurred within a few tens of years. 

Since we resolve the knots within the blue- and red-shifted jet, we also estimated the dynamical timescale of the knots. 
For the furthest northern (blue-shifted) knots (N15), we estimated a dynamical age of 191.2 yr whilst the dynamical age of the furthest southern (red-shifted) knot (S8) has a dynamical age of 95.8 yr. 
Although the jet extent is much longer to the north, the dynamical age is much smaller than what we derived for the outflow. 
In fact, if we consider the dynamical age of the knot N6, whose position corresponds to where the \textit{arc-like} feature appears and which is at the level of the shock positions B and E, then it is much smaller ($\tau_{\text{dyn}_N6}\sim 124$ yr) than that of the outflow. A caveat in our derived dynamical timescale is associated with the uncertainties resulting from the assumed typical jet velocity of 100 \kms (Section~\ref{subsec:tdyn}). As discussed in \cite{podio2021}, taking into account all the uncertainties associated with the estimates of de-projected velocities, the typical protostellar jet velocities are always consistent with $V_{\mathrm{jet}}=100 \pm 50$ \kms. Hence, the inferred $\tau_{\mathrm{dyn}}$ may be a factor 2 larger or 1.5 times smaller, if the jet velocity is 50 \kms or 150 \kms, respectively. However, the dynamical scale would still be below the age derived by our chemical model for the outflow.
Therefore, it seems that the outflow cavity and the jets do not have the same dynamic timescale. Differences between the outflow and jet dynamical timescales have already been measured towards other sources \citep[e.g.][]{hirano2010, matsushita2019, takahashi2024}.
This result is also coherent with recent hydro-dynamical simulations of molecular jet-driven outflows that showed that the dynamical timescale of the jet is lower than the duration of the ejection phase \citep{rivera-ortiz2023}.

In Sec.~\ref{subsec: collimation}, we found that the northern part of the jet could be broken down into two jet events, with the younger one terminating at the \textit{arc-like} feature, whilst the older one is seen beyond it. 
The ejection process in outflows is known to create jets, bow-shocks and cavities \citep[e.g.][]{bally2016, rivera-ortiz2023}. 
Therefore, assuming that the two jets delimited by the \textit{arc-like} feature correspond indeed to different ejection events, then the \textit{arc-like} feature is actually another bow shock within the northern part of the outflow. 
This would be consistent with the shocked gas traced by \ce{CH3OH} at positions B and E, as well as the presence of \ce{CH3OH} masers. 
Additional support to this scenario comes from the dynamical time interval of the blue-shifted knots in Fig.~\ref{fig:jet-tdyn-interv}, where we can see a clear difference in behaviour before and after 2000 au, which allegedly marks the location of the most recent bow shock. 
Whilst up to $\sim$ 1800 au the interval between the knots increases up to $\sim20$ yr, it suddenly decreases at 2000 au with $\Delta_{\tau_{dyn}}\sim 10$ yr and varies without clear trend at larger distances ($> 2000 au$). 
We interpret this with the fact that the first jet (i.e. the older event) encountered more material and was, thus, slowed down, expanding in a relatively dense environment. 
Once the second event was launched, the jet was thus propagating in a \textit{cleared} medium, thanks to the first event clearing away the dense material of the cloud and, as it is not slowed down, it catches up with the older slower jet material, thus leading to the formation of the second recent bow-shock, which is the \textit{arc-like} feature \citep[e.g.][]{bally2016}. On the other hand, the two different timescales seen in the knot interval distribution could also reflect different ejection modes with different periodicity \citep[e.g.][]{raga1990, rabenanahary2022, rivera-ortiz2023, lora2024}.

Finally, multiple outflow events have been observed towards other sources, such as HOPS373-SW  \citep{lee2024}, IRAS 15398-3359 \citep{okoda2021} or the IRAS 4A system \citep{chahine2024}.
However, whilst in these cases the outflow events underwent a reorientation or precession, for example caused by the presence of a binary system like in the case of HOPS 373-SW \citep{lee2024} or IRAS4A \citep{chahine2024}, it does not seem to be the case for FIR6c-a, where the two bow-shocks seem to be relatively aligned with each other.

\subsection{A highly CR-irradiated system}

Figure \ref{fig:model-CH3OH-CH3CN} shows the comparison between model predictions and observations. 
Remarkably, the predicted [\ce{CH3OH}]/[\ce{CH3CN}] ratio reproduces the observation in both the outflow and hot corino positions only for a large CR ionization rate (CRIR), $\zeta_{CR}=1\times10^{-14}$, much larger than the assumed average galactic value of $\zeta_0=3\times 10^{-17}$ \s \citep[e.g.][]{padovani2009a}. 
Such a high value could be explained by the fact that CR can be locally accelerated in shocks located in protostellar jets or the inner region of the protostellar envelopes and disks \citep{padovani2015, padovani2016, GO18}. 

Large $\zeta_{CR}$ values have previously been evoked in the OMC-2/FIR4 protocluster, which is located to the north of FIR6c-a within the OMC-2/3 filament \citep{ceccarelli2014, fontani2017, favre2018}. 
Specifically, the FIR4 envelope seems to be permeated by a flux of ionizing particles about 1000 times larger than the average interstellar value, i.e., similar to the one derived from our chemical modelling in FIR6c-a. 
In the case of FIR4, it has been proposed that the CR-like particles (with E $\geq$ 10 MeV) originate in the inner part of the FIR4 envelope, from a young protostar undergoing energetic eruptions \citep{ceccarelli2014}. 
Alternatively, \citep{sato2023} proposed that the high derived $\zeta_{CR}$ is rather due to the interaction between the outflow driven by FIR3 with the FIR4 envelope. 
A more recent study by \cite{lattanzi2023} suggested that the jet driven by the Class 0 protostar HOPS 108, one of the sources located within the FIR4 cluster, could be at the origin of the local enhancement of CR in the region.

It thus seems that FIR6 and its outflow are highly CR-irradiated, a characteristic which might be common to all the OMC-2/3 filament. 
However, are the clusters FIR4 and FIR6 unique within the OMC-2/3 filament or is the whole OMC-2/3 region presenting a high CRIR? 
A theoretical study by \cite{GO18} has shown that large protoclusters (with $N_* \geq $ a few hundred), such as the OMC filament, accelerate CR and provide a larger CR ($\zeta > 10^{-16}$ \s) within their natal cloud. 
On the observational side, \cite{socci2024} measured the CRIR at parcsec scales towards the OMC-2 and 3 clouds. 
They found that the two regions show different values of CRIR, with the highest values ($\zeta \gtrsim 2\times 10^{-16}$ \s) towards the OMC-2 region. 
The protostars located in the OMC-2 region, therefore, could have formed in a naturally CR-enriched environment. 

On the other hand, high CRIR could not be related to the specific environment provided by the OMC-2/3 filament but rather to the star formation process itself. 
High CRIRs were also found in other sources outside the Orion star-forming region. 
Indeed, \cite{podio2014} found $\zeta=3\times10^{-16}$ \s towards the shock position B1 within the L1157-mm outflow.  
More recently,  CR enhancements were also found towards the B335 protostar ($\zeta\sim 10^{-14}$ \s at a few hundred au from the central object; \citealt{cabedo2023}) and in the vicinity of protostars located within the NGC 1333 cloud ($\zeta\sim 10^{-16.5}$ \s; \citealt{pineda2024}).
Larger scale (parsec) studies of high-mass star-forming regions, including OMC-2/3, have shown that the CRIR can vary from one environment to another. 
Studies towards other sources located in the OMC-2/3 filament are needed to understand what causes such high elevated CRIR towards the FIR6c-a source.

%------------------------------------------------------------

\section{Conclusions}\label{sec:conclusions}
We report the study of the outflow driven by the FIR6c-a source within the OMC-2/3 filament, using high sensitivity and angular resolution (0.25$\arcsec$ or $\sim 100$ au) ALMA observations, obtained within the ORANGES project. 
The major conclusions can be summarized in the following items.

(1) In addition to SiO, we detect emission from \ce{CH3OH} and \ce{CH3CN}, for the first time towards the outflow, which makes the OMC-2 FIR6c-a outflow the second iCOM-rich outflow detected in the OMC-2/3 filament after the one emanating from HOPS-87 \citep{hsu2024}, and the second outflow ever mapped in \ce{CH3CN} after L1157-B1 \citep{codella2009}. 
This new chemically active outflow, similar in shape and extension to the well-known L1157 outflow, will allow comparative detailed studies of the impact of the environment on the chemical appearance of outflows.

(2) \ce{CH3OH} and \ce{CH3CN} are only detected in the northern lobe of the FIR6c-a outflow, which suggests the presence of a medium denser in the north with respect to the south.

(3) While SiO is observed both in the jet and the outflow cavity walls, methanol is detected along the cavity walls and in two arc-like features joining the cavity walls, one where the SiO jet emission ends and the second one midway ($\sim 2000$ au from the central source). 
Methyl cyanide emission is only detected in the midway arc.

(4) The high sensitivity and angular resolution of the ALMA observations allowed us to detect the red-shifted counterpart of the jet, again for the first time, and to resolve the knotty structure of the blue- and red-shifted jets. 
The SiO northern jet is much more extended (9400 au) than the southern one, only visible within 2300 au from the central source.
It is also less collimated ($\sim 5^\circ$) than the southern SiO jet ($\leq 1.5^\circ$).
Using the SiO emission, we derived the dynamical times of the northern and southern jets' knots, which are, on the contrary to the other properties, relatively similar, with a break at $\geq$2000 au from the central source.

(5) We found that two blue-shifted jets issued from two different outflow events can naturally explain the arc-like features traced by \ce{CH3OH} and \ce{CH3CN}, as well as the observed \say{break down} of the dynamical times of the knots in the jets. 

(6) The comparison of the measured abundance ratio [\ce{CH3OH}]/[\ce{CH3CN}] with astrochemical model predictions allowed us to constrain the age of the outflow, $\geq 1000 $ yr, and the CR ionisation rate,  $10^{-14}$ \s. 
It is not clear whether the high CR ionisation rate is due to the outflow itself and is, thus, related to the star formation process, or whether the OMC-2/3 filament is naturally enhanced in CR because of the presence of several forming stars. 
Further studies towards other sources within the filament are needed to distinguish between the two scenarios.

\section*{Acknowledgements}
We thank the anonymous referee for their comments and suggestions, which helped improve the paper.
This project has received funding from the European Research Council (ERC) under the European Union’s Horizon 2020 research and innovation
program, for the Project “The Dawn of Organic Chemistry” (DOC),
grant agreement No 741002. M.B acknowledges the support from the European Research Council (ERC) Advanced grant MOPPEX 833460.  This paper makes use of the following ALMA data: ADS/JAO.ALMA\#2016.1.00376.S. ALMA is a partnership of ESO (representing its member states), NSF (USA) and NINS (Japan), together with NRC (Canada) and NSC and ASIAA (Taiwan), in cooperation with the Republic of Chile. The Joint ALMA Observatory is operated by ESO, AUI/NRAO and NAOJ.
%%%%%%%%%%%%%%%%%%%%%%%%%%%%%%%%%%%%%%%%%%%%%%%%%%
\section*{Data Availability}
The raw data are available on the ALMA archive (ADS/JAO.ALMA\#2016.1.00376.S).
 
%The inclusion of a Data Availability Statement is a requirement for articles published in MNRAS. Data Availability Statements provide a standardised format for readers to understand the availability of data underlying the research results described in the article. The statement may refer to original data generated in the course of the study or to third-party data analysed in the article. The statement should describe and provide means of access, where possible, by linking to the data or providing the required accession numbers for the relevant databases or DOIs.

%%%%%%%%%%%%%%%%%%%% REFERENCES %%%%%%%%%%%%%%%%%%

% The best way to enter references is to use BibTeX:

\bibliographystyle{mnras}
\bibliography{outflow_main.bib} % if your bibtex file is called example.bib

% Alternatively you could enter them by hand, like this:
% This method is tedious and prone to error if you have lots of references
%\begin{thebibliography}{99}
%\bibitem[\protect\citeauthoryear{Author}{2012}]{Author2012}
%Author A.~N., 2013, Journal of Improbable Astronomy, 1, 1
%\bibitem[\protect\citeauthoryear{Others}{2013}]{Others2013}
%Others S., 2012, Journal of Interesting Stuff, 17, 198
%\end{thebibliography}

%%%%%%%%%%%%%%%%%%%%%%%%%%%%%%%%%%%%%%%%%%%%%%%%%%

%%%%%%%%%%%%%%%%% APPENDICES %%%%%%%%%%%%%%%%%%%%%

\appendix

\section{Moment maps}\label{app:moment}

\begin{figure*}
    \centering
    \includegraphics[width=0.7\textwidth]{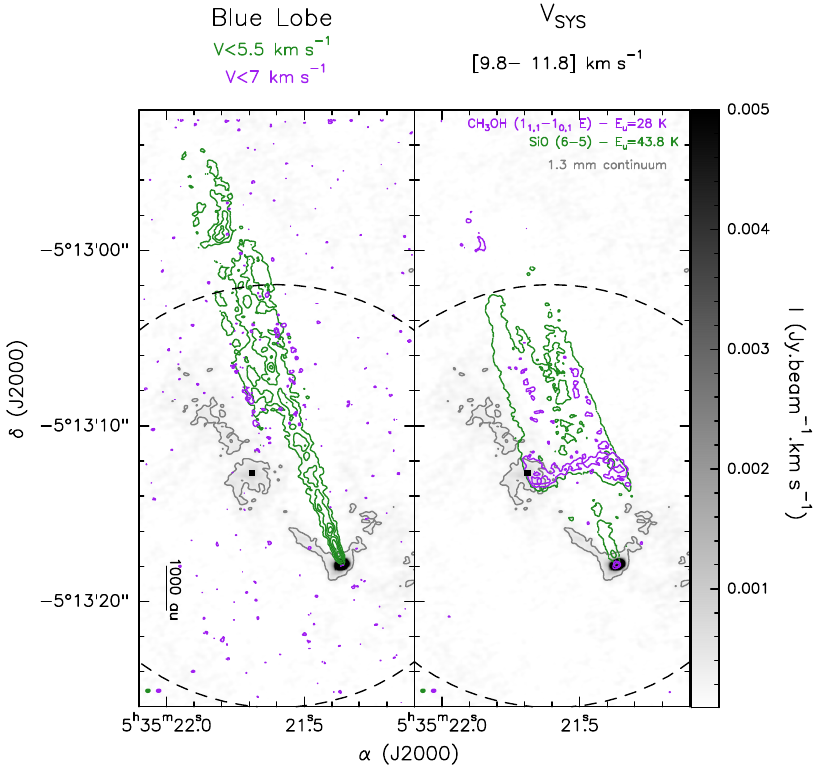}
    \caption{Moment 0 maps of SiO (6-5; green contours) and \ce{CH3OH} ($1_{1,1}-1_{0,1}$ E; purple contours), overlaid on the 1.3mm continuum emission (grey-scaled background). For the 1.3mm continuum, only the 3$\sigma$ level is shown (grey contour; 1$\sigma=60\mu$Jy.beam$^{-1}$\kms.) The moment maps are shown at three different velocity components: blue-shifted ($V<5.5$ and $V<7$ \kms for SiO and \ce{CH3OH}, respectively; \textit{left}), around the systemic velocity ([$10.8 \pm 1$] \kms; \textit{right}), respectively. Levels of SiO start at 10 and 5$\sigma$ ($1\sigma=2.6$ mJy.beam$^{-1}$\kms) with steps of 40 and 15$\sigma$, respectively. Levels of \ce{CH3OH} start at 3 $\sigma$ ($1\sigma=2.7$ mJy.beam$^{-1}$\kms) with steps of 3$\sigma$ (\textit{left}) and 5$\sigma$ (\textit{right}). The black filled square mark the position of the starless core FIR6c-c. The dashed black circle represents the primary beam of 24.1$\arcsec$ from the \ce{CH3OH} observations. The beams of both the SiO and \ce{CH3OH} observations are depicted in the bottom left corner of each panel.}
    \label{fig:ch3oh-28}
\end{figure*}

\begin{figure*}
    \centering
    \includegraphics[width=0.7\textwidth]{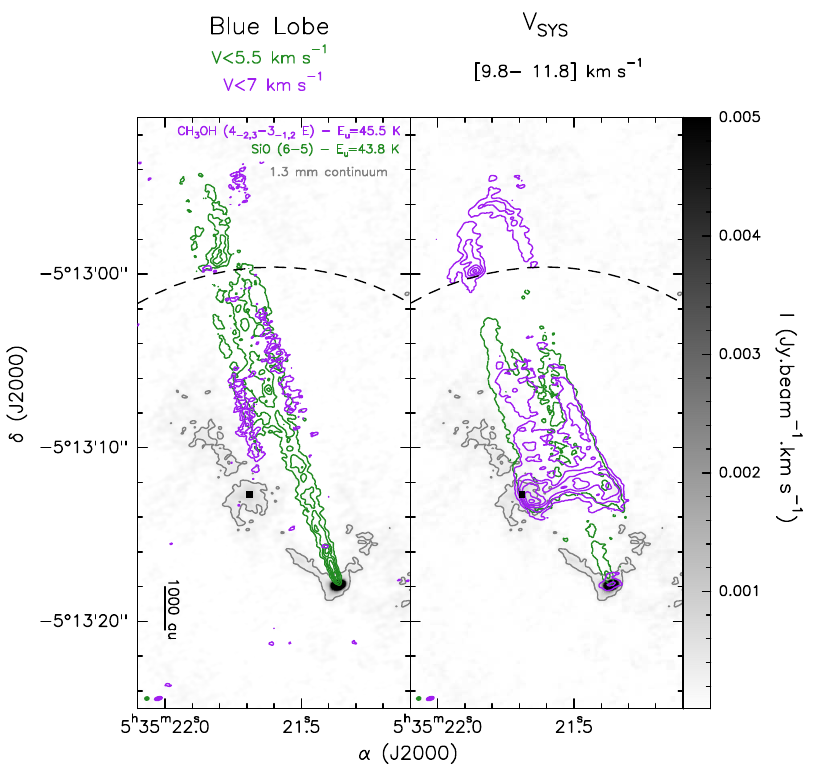}
    \caption{Moment 0 maps of SiO (6-5; green contours) and \ce{CH3OH} ($4_{-2,3}-1_{-1,2}$ E; purple contours), overlaid on the 1.3mm continuum emission (grey-scaled background). For the 1.3mm continuum, only the 3$\sigma$ level is shown (grey contour; 1$\sigma=60\mu$Jy.beam$^{-1}$\kms.) The moment maps are shown at three different velocity components: blue-shifted ($V<5.5$ and $V<7$ \kms for SiO and \ce{CH3OH}, respectively; \textit{left}), around the systemic velocity ([$10.8 \pm 1$] \kms; \textit{right}), respectively. Levels of SiO start at 10 and 5$\sigma$ ($1\sigma=2.6$ mJy.beam$^{-1}$\kms) with steps of 40 and 15$\sigma$, respectively. Levels of \ce{CH3OH} start at 5$\sigma$ ($1\sigma=3.5$ mJy.beam$^{-1}$\kms) with steps of 4$\sigma$ (\textit{left}) and 8$\sigma$ (\textit{right}). The black filled square mark the position of the starless core FIR6c-c. The dashed black circle represents the primary beam of 28.8$\arcsec$ from the \ce{CH3OH} observations. The beams of both the SiO and \ce{CH3OH} observations are depicted in the bottom left corner of each panel.}
    \label{fig:ch3oh-45}
\end{figure*}

Figures \ref{fig:ch3oh-28} and \ref{fig:ch3oh-45} show the moment 0 maps of the two other transitions of \ce{CH3OH} at 28 K and 45.5 K, respectively. The transition at 28K is very likely suffering from filtered emission.

%%%%%%%%%%%%%%%%%%%%%%

\section{Spectra and Gaussian Fits}\label{app:spectra}
\begin{figure*}
    \centering
    \includegraphics[width=0.7\linewidth]{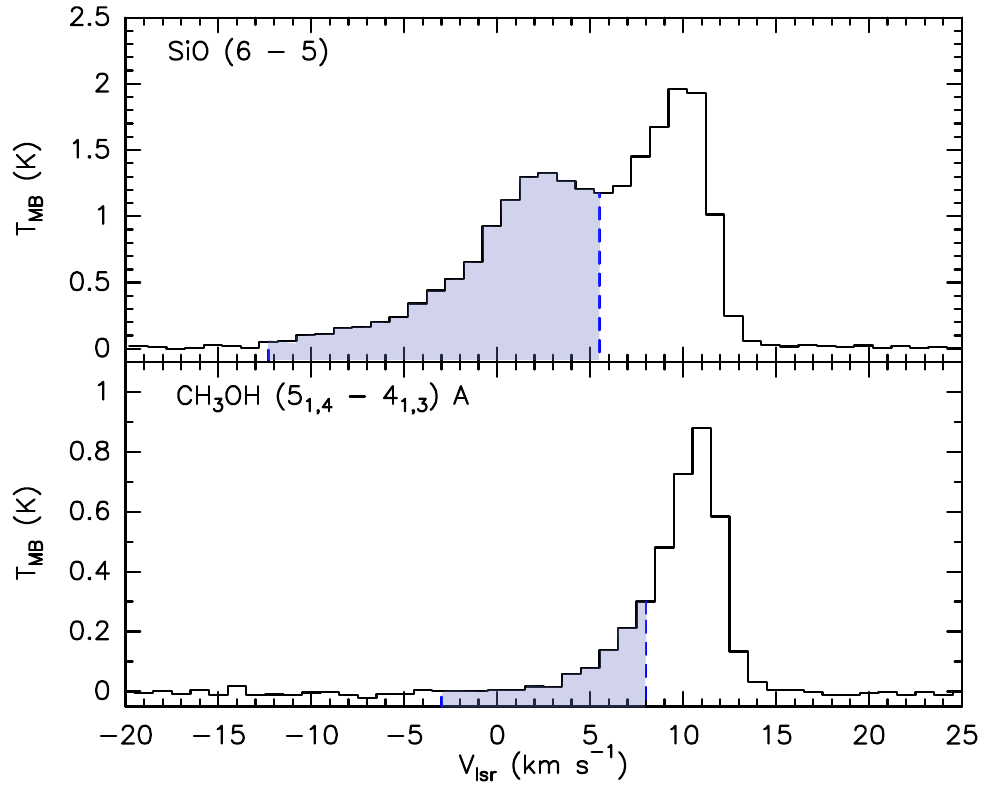}
\caption{Line profiles for SiO (6-5) and \ce{CH3OH}($5_{1,4}-4_{1,3}$) A integrated over the full emission of the two species, i.e. in the range [-12.3 ; +26.7]\kms and [-5.5; +11.8] \kms, respectively. The range of velocity used to map the blue-shifted component of the outflow in Fig.~\ref{fig:moment-maps} is highlighted in blue and delimited by vertical blue dashed lines. The source's $V_{\text{sys}}$ is 10.8 \kms.}
\label{fig:full_lines-profiles}
\end{figure*}

The full lines profiles for SiO and \ce{CH3OH} towards the northern part of the jet are shown in Fig.~\ref{fig:full_lines-profiles}.
The spectra extracted from a $0.5\arcsec\times0.5\arcsec$ region towards position A (only for \ce{CH3CN}) to E are shown below in Figures~\ref{Fig: spectra_HC} to \ref{Fig: spectra_WCS}. The results of the Gaussian fit are presented in Table~\ref{tab:fits}. We note that the uncertainty reported in Table~\ref{tab:fits} are obtained from the Gaussian fitting. However, the spectral resolution is 0.5 \kms, hence, the differences in the peak velocities and line widths of \ce{CH3CN} for position B are not significant to consider the presence of two components. Both the emission of \ce{CH3OH} and \ce{CH3CN} are thus associated with the outflow.

\begin{figure*}
    \centering
    \includegraphics[width=0.7\linewidth]{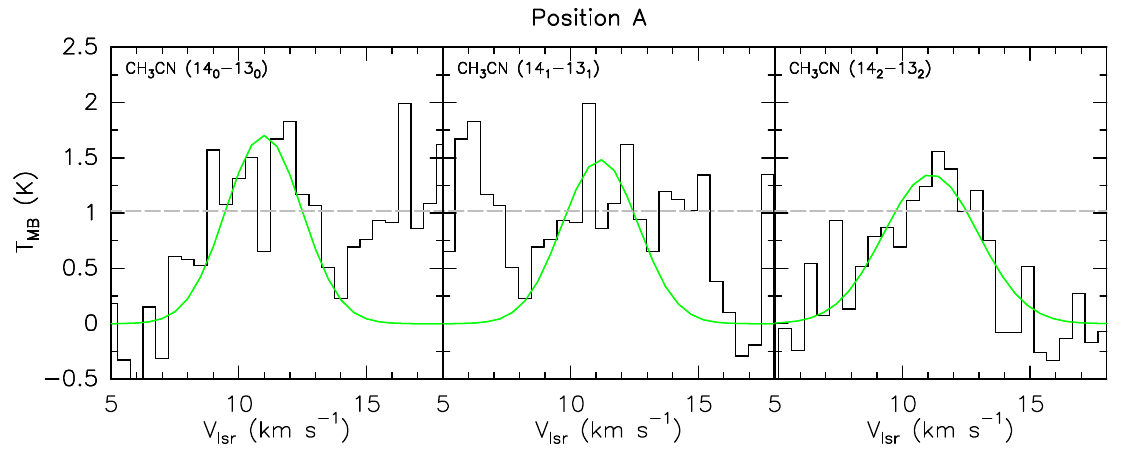}
\caption{Spectra of \ce{CH3CN} towards position A (corresponding to the hot corino. The solid green line shows the Gaussian fit performed and the dashed grey line indicates a $3\sigma$ level. }
    \label{Fig: spectra_HC}
\end{figure*}

\begin{figure*}
    \centering
    \includegraphics[width=0.7\linewidth]{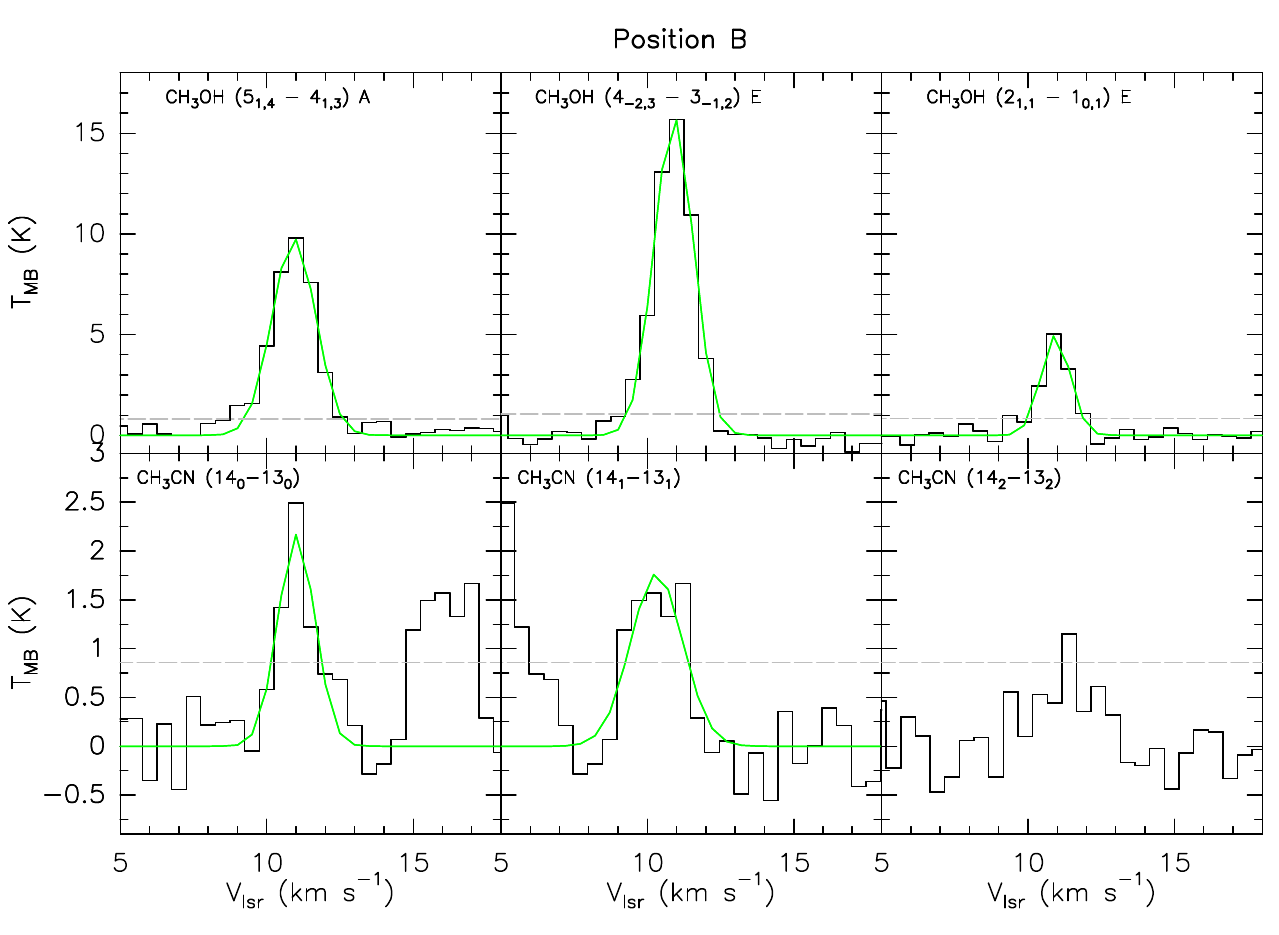}
    \caption{Spectra of \ce{CH3OH} and \ce{CH3CN} towards Position B. The solid green line shows the Gaussian fit performed and the dashed grey line indicate a 3$\sigma$ level.}
    \label{Fig: spectra_ECS}
\end{figure*}

\begin{figure*}
    \centering
    \includegraphics[width=0.7\linewidth]{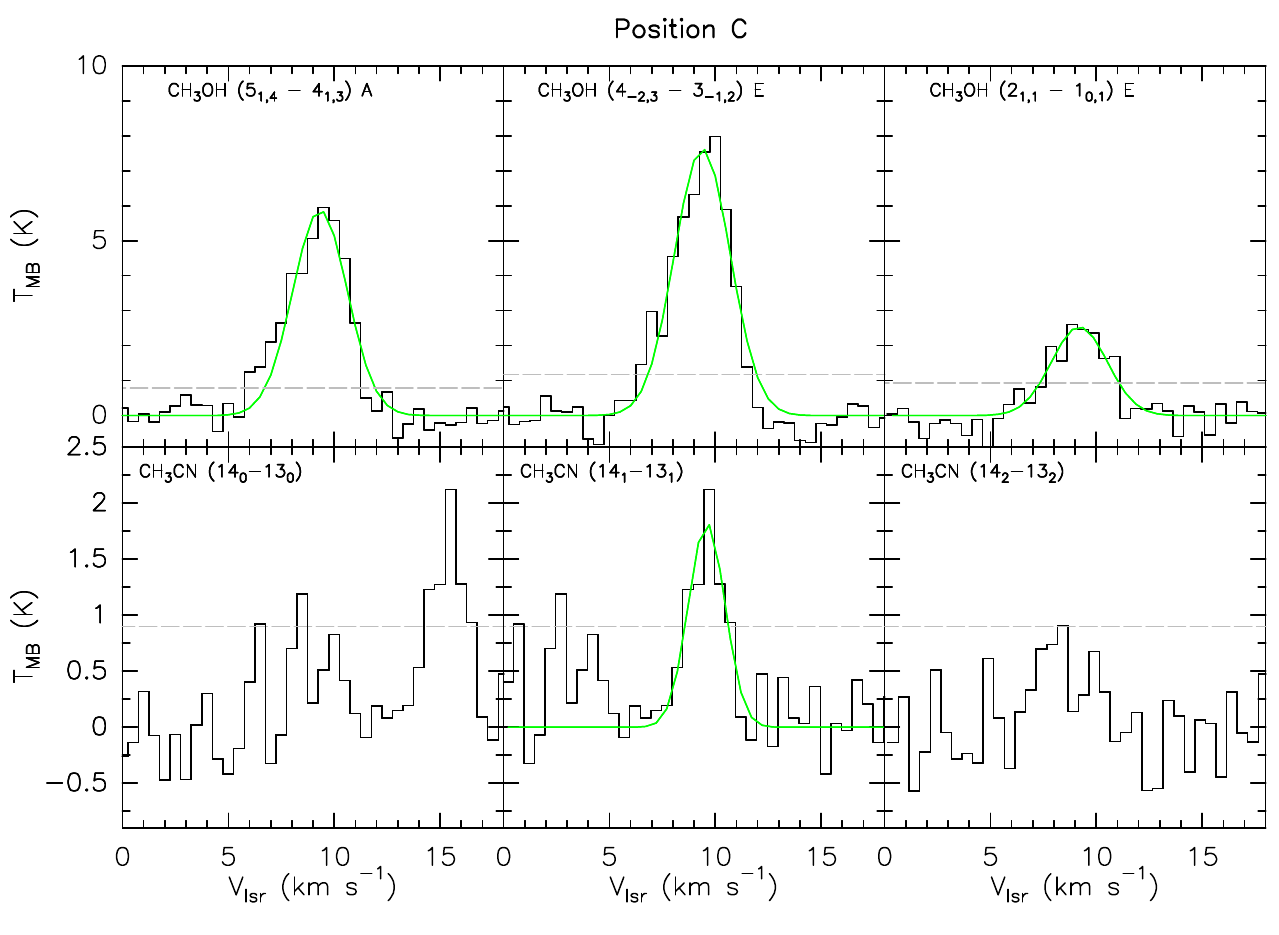}
    \caption{Spectra of \ce{CH3OH} and \ce{CH3CN} towards Position C. The solid green line shows the Gaussian fit performed and the dashed grey line indicate a 3$\sigma$ level.}
    \label{Fig: spectra_ECW}
\end{figure*}

\begin{figure*}
    \centering
    \includegraphics[width=0.7\linewidth]{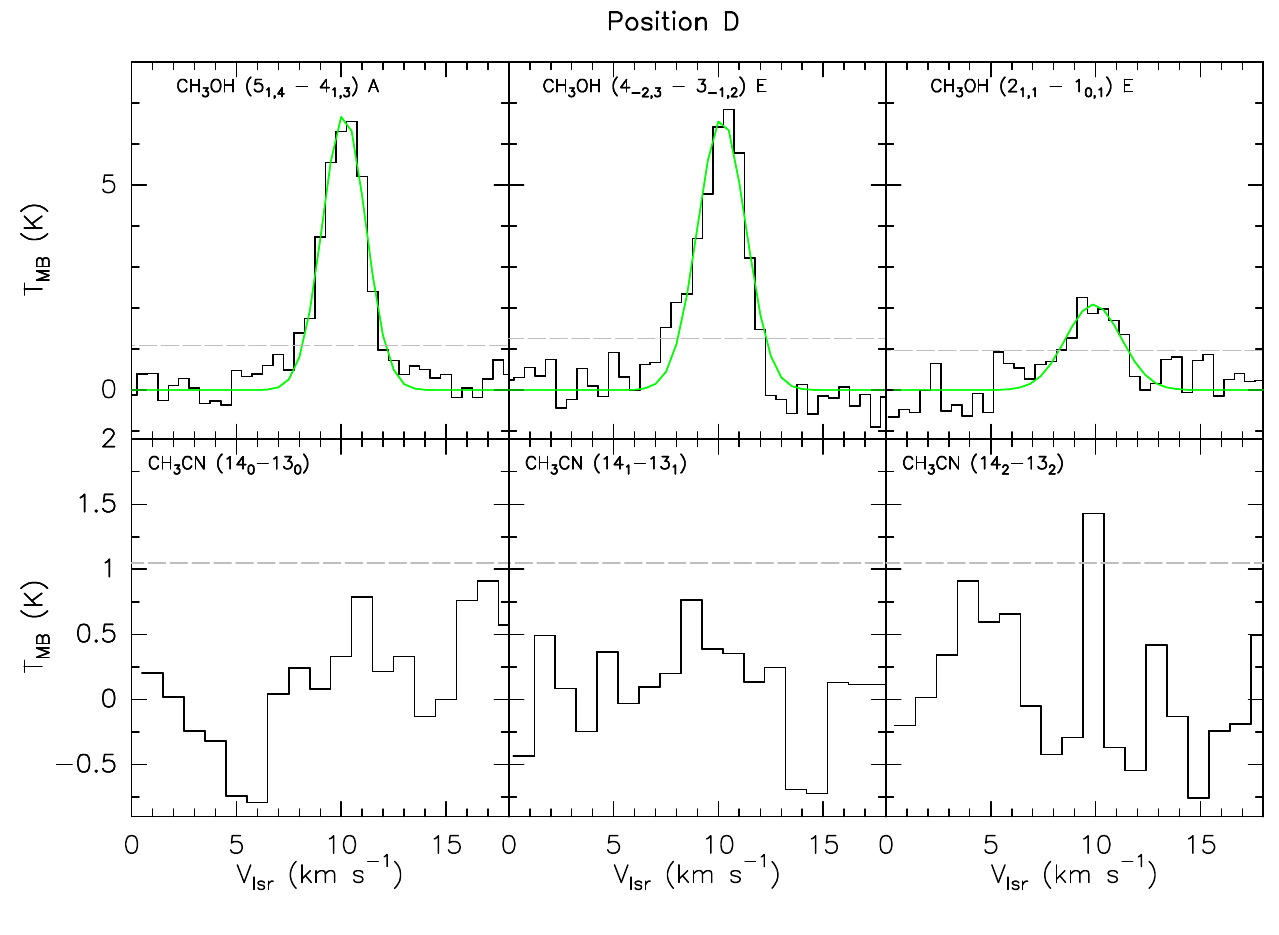}
    \caption{Spectra of \ce{CH3OH} and \ce{CH3CN} towards Position D. The solid green line shows the Gaussian fit performed and the dashed grey line indicate a 3$\sigma$ level.}
    \label{Fig: spectra_WCW}
\end{figure*}

\begin{figure*}
    \centering
    \includegraphics[width=0.7\linewidth]{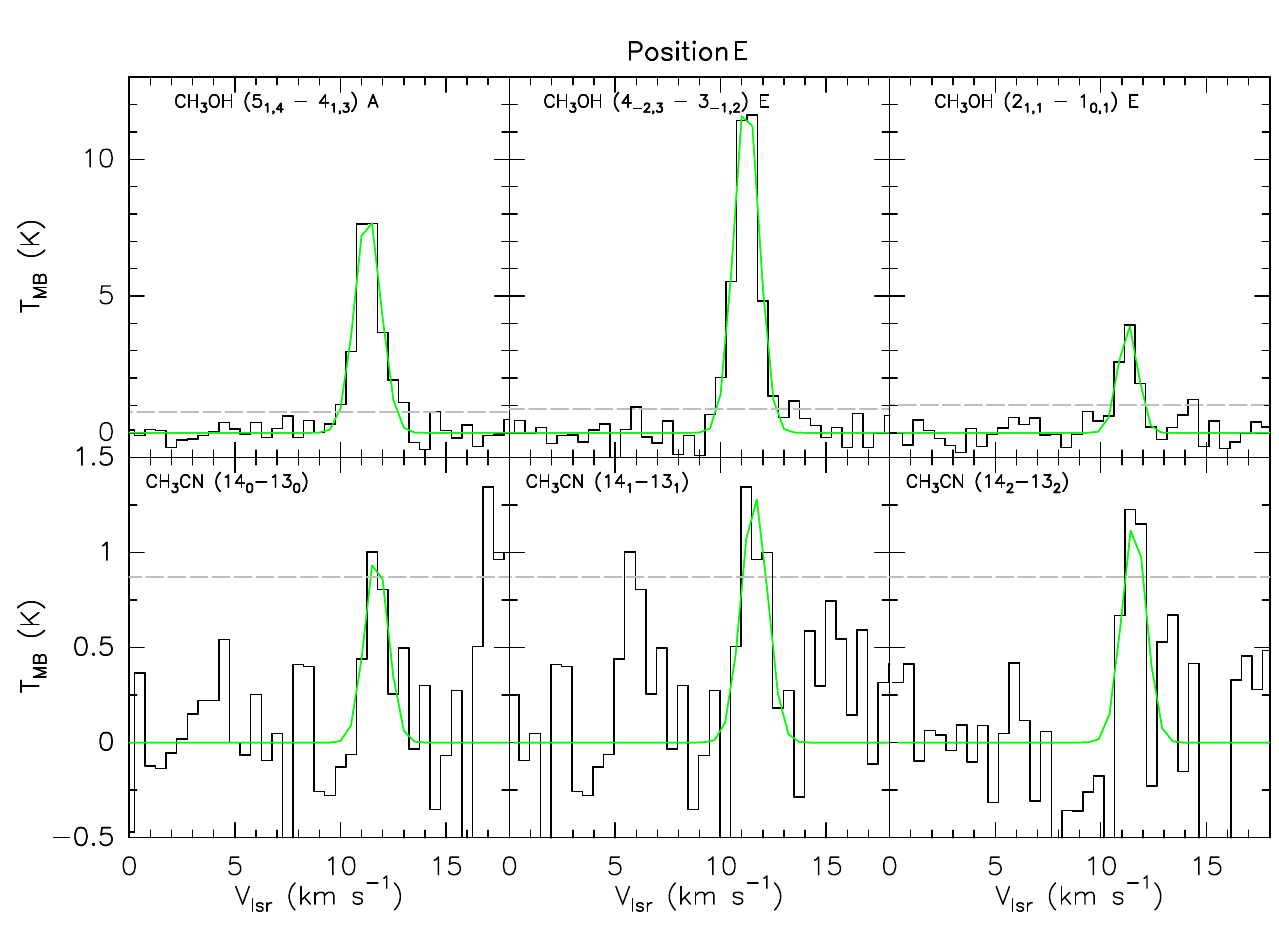}
    \caption{Spectra of \ce{CH3OH} and \ce{CH3CN} towards Position E. The solid green line shows the Gaussian fit performed and the dashed grey line indicate a 3$\sigma$ level.}
    \label{Fig: spectra_WCS}
\end{figure*}

\begin{table*}
    \centering
    \caption{Gaussian Line Fitting Results. The calibration uncertainty of 10\% is not included in the integrated intensity error and the spectral resolution is 0.5 \kms.}
    \label{tab:fits}
    \begin{tabular}{cccccc}
    \hline \hline
        Species & Rest Frequency & $\int T_{\text{MB}}dV$ & $V_{\text{peak}}$& FWHM & rms  \\
         &(MHz) & (K.\kms) & (\kms) & (\kms)& (K) \\
         \hline
         \multicolumn{6}{c}{Position A}\\
         \hline
         \multirow{3}{*}{\ce{CH3CN}} &257507 &6.25 $\pm$ 0.85 &11.1 $\pm$ 0.3 & 4.3 $\pm$ 0.7 & \multirow{3}{*}{0.34} \\
         &257522 &5.52 $\pm$ 0.63 & 11.2 $\pm$ 0.3 & 3.5 $\pm$ 0.5 &  \\
         &257527 &6.34 $\pm$ 0.63 &11.0 $\pm$ 0.3 & 3.5 $\pm$ 0.5 &\\
         \hline
          \multicolumn{6}{c}{Position B}\\
          \hline
         \multirow{3}{*}{\ce{CH3OH}} &218440 &26.50  $\pm$ 0.51 &10.9 $\pm$ 0.1 & 1.6 $\pm$ 0.1 & 0.35 \\
         & 243915 &18.28 $\pm$ 0.45 & 10.9 $\pm$ 0.1 & 1.8 $\pm$ 0.1& 0.27 \\
         &261805 &6.21 $\pm$ 0.39 & 10.9 $\pm$ 0.1 & 1.2 $\pm$ 0.1& 0.28\\
         \hline
         \multirow{2}{*}{\ce{CH3CN}} &257522 &3.76 $\pm$ 0.40 &10.3 $\pm$ 0.1 & 2.1 $\pm$ 0.3 & \multirow{2}{*}{0.31} \\
         &257527 &3.61 $\pm$ 0.49 & 11.0 $\pm$ 0.1 & 1.5 $\pm$ 0.3 &\\
         \hline
         \multicolumn{6}{c}{Position C}\\
          \hline
         \multirow{3}{*}{\ce{CH3OH}} &218440 & 25.26 $\pm$ 0.82 & 9.4 $\pm$ 0.1& 3.1 $\pm$ 0.1&0.39\\
         & 243915 &19.06 $\pm$ 0.59&9.3 $\pm$ 0.1&3.0 $\pm$ 0.1&0.26 \\
         &261805 &8.38 $\pm$ 0.61&9.2 $\pm$ 0.1&3.1 $\pm$ 0.3&0.31\\
         \hline
         \multirow{1}{*}{\ce{CH3CN}} &257522 & 3.92 $\pm$ 0.48& 9.6 $\pm$ 0.1 & 2.0 $\pm$ 0.3&0.30 \\
         \hline
         \multicolumn{6}{c}{Position D}\\
          \hline
         \multirow{3}{*}{\ce{CH3OH}} &218440 &19.02 $\pm$ 0.77 & 10.2 $\pm$ 0.1 & 2.7 $\pm$ 0.1 &0.42 \\
         & 243915 &17.44 $\pm$ 0.61 & 10.1 $\pm$ 0.1&2.4 $\pm$ 0.1&0.36 \\
         &261805 &6.92 $\pm$ 0.91& 9.9 $\pm$ 0.2 &3.1 $\pm$ 0.5&0.32\\
         \hline
         \multicolumn{6}{c}{Position E}\\
          \hline
         \multirow{3}{*}{\ce{CH3OH}} &218440 &18.38 $\pm$ 0.46 & 11.2 $\pm$ 0.1&1.4 $\pm$ 0.1&0.29 \\
         & 243915 &12.48 $\pm$ 0.39&11.3 $\pm$ 0.1&1.4 $\pm$ 0.1&0.25 \\
         &261805 &4.62 $\pm$ 0.38&11.3 $\pm$ 0.1& 1.1 $\pm$ 0.1&0.34\\
         \hline
         \multirow{3}{*}{\ce{CH3CN}} &257507 & 1.31 $\pm$ 0.35&11.6 $\pm$ 0.1&0.9 $\pm$ 0.2&\multirow{3}{*}{0.33}\\
         &257522 &1.76 $\pm$ 0.41&11.6 $\pm$ 0.2&1.4 $\pm$ 0.3&\\
         &257527 &1.09 $\pm$ 0.40&11.7 $\pm$ 0.2&1.1 $\pm$ 0.5&\\
         \hline
    \end{tabular}
\end{table*}
%%%%%%%%%%%%%%%%%%%%%%%%%%%%%%%%%%%%%%%%%%%%%%%%%%

% Don't change these lines
\bsp	% typesetting comment
\label{lastpage}
\end{document}